\documentclass[amsmath,amssymb,aps,array,nofootinbib,nolongbibliography,superscriptaddress]{revtex4-2}

\usepackage{bm}
\usepackage{dcolumn}
\usepackage{graphicx}
\usepackage{multirow}
\usepackage{color}
\usepackage{mathtools}
\usepackage[T1]{fontenc}
\usepackage{lmodern}

\newcommand{\PreserveBackslash}[1]{\let\temp=\\#1\let\\=\temp}
\newcolumntype{C}[1]{>{\PreserveBackslash\centering}p{#1}}
\newcolumntype{R}[1]{>{\PreserveBackslash\raggedleft}p{#1}}
\newcolumntype{L}[1]{>{\PreserveBackslash\raggedright}p{#1}}

\begin{document}

\title{Searching non-standard interactions with atmospheric neutrinos at ESSnuSB}

\author{J.~Aguilar}
\affiliation{Consorcio ESS-bilbao, Parque Cient\'{i}fico y Tecnol\'{o}gico de Bizkaia, Laida Bidea, Edificio 207-B, 48160 Derio, Bizkaia, Spain}
\author{M.~Anastasopoulos}
\affiliation{European Spallation Source, Box 176, SE-221 00 Lund, Sweden}
\author{D.~Bar\u{c}ot}
\affiliation{Center of Excellence for Advanced Materials and Sensing Devices, Ruđer Bo\v{s}kovi\'c Institute, 10000 Zagreb, Croatia}
\author{E.~Baussan}
\affiliation{IPHC, Universit\'{e} de Strasbourg, CNRS/IN2P3, Strasbourg, France}
\author{A.K.~Bhattacharyya}
\affiliation{European Spallation Source, Box 176, SE-221 00 Lund, Sweden}
\author{A.~Bignami}
\affiliation{European Spallation Source, Box 176, SE-221 00 Lund, Sweden}
\author{M.~Blennow}
\affiliation{Department of Physics, School of Engineering Sciences, KTH Royal Institute of Technology, Roslagstullsbacken 21, 106 91 Stockholm, Sweden}
\affiliation{The Oskar Klein Centre, AlbaNova University Center, Roslagstullsbacken 21, 106 91 Stockholm, Sweden}
\author{M.~Bogomilov}
\affiliation{Sofia University St. Kliment Ohridski, Faculty of Physics, 1164 Sofia, Bulgaria}
\author{B.~Bolling}
\affiliation{European Spallation Source, Box 176, SE-221 00 Lund, Sweden}
\author{E.~Bouquerel}
\affiliation{IPHC, Universit\'{e} de Strasbourg, CNRS/IN2P3, Strasbourg, France}
\author{F.~Bramati}
\affiliation{University of Milano-Bicocca and INFN Sez. di Milano-Bicocca, 20126 Milano, Italy}
\author{A.~Branca}
\affiliation{University of Milano-Bicocca and INFN Sez. di Milano-Bicocca, 20126 Milano, Italy}
\author{G.~Brunetti}
\affiliation{University of Milano-Bicocca and INFN Sez. di Milano-Bicocca, 20126 Milano, Italy}
\author{I.~Bustinduy}
\affiliation{Consorcio ESS-bilbao, Parque Cient\'{i}fico y Tecnol\'{o}gico de Bizkaia, Laida Bidea, Edificio 207-B, 48160 Derio, Bizkaia, Spain}
\author{C.J.~Carlile}
\affiliation{Department of Physics, Lund University, P.O Box 118, 221 00 Lund, Sweden}
\author{J.~Cederkall}
\affiliation{Department of Physics, Lund University, P.O Box 118, 221 00 Lund, Sweden}
\author{T.~W.~Choi}
\affiliation{Department of Physics and Astronomy, FREIA Division, Uppsala University, P.O. Box 516, 751 20 Uppsala, Sweden}
\author{S.~Choubey}
\email{Corresponding authors: S. Choubey (choubey@kth.se), T. Ohlsson (tohlsson@kth.se) and S. Vihonen (vihonen@kth.se)}
\affiliation{Department of Physics, School of Engineering Sciences, KTH Royal Institute of Technology, Roslagstullsbacken 21, 106 91 Stockholm, Sweden}
\affiliation{The Oskar Klein Centre, AlbaNova University Center, Roslagstullsbacken 21, 106 91 Stockholm, Sweden}
\author{P.~Christiansen}
\affiliation{Department of Physics, Lund University, P.O Box 118, 221 00 Lund, Sweden}
\author{M.~Collins}
\affiliation{Faculty of Engineering, Lund University, P.O Box 118, 221 00 Lund, Sweden}
\affiliation{European Spallation Source, Box 176, SE-221 00 Lund, Sweden}
\author{E.~Cristaldo Morales}
\affiliation{University of Milano-Bicocca and INFN Sez. di Milano-Bicocca, 20126 Milano, Italy}
\author{P.~Cupia\l}
\affiliation{AGH University of Krakow, al. A. Mickiewicza 30, 30-059 Krakow, Poland }
\author{D.~D'Ago}
\affiliation{INFN Sez. di Padova, Padova, Italy}
\author{H.~Danared}
\affiliation{European Spallation Source, Box 176, SE-221 00 Lund, Sweden}
\author{J.~P.~A.~M.~de~Andr\'{e}}
\author{M.~Dracos}
\affiliation{IPHC, Universit\'{e} de Strasbourg, CNRS/IN2P3, Strasbourg, France}
\author{I.~Efthymiopoulos}
\affiliation{CERN, 1211 Geneva 23, Switzerland}
\author{T.~Ekel\"{o}f}
\affiliation{Department of Physics and Astronomy, FREIA Division, Uppsala University, P.O. Box 516, 751 20 Uppsala, Sweden}
\author{M.~Eshraqi}
\affiliation{European Spallation Source, Box 176, SE-221 00 Lund, Sweden}
\author{G.~Fanourakis}
\affiliation{Institute of Nuclear and Particle Physics, NCSR Demokritos, Neapoleos 27, 15341 Agia Paraskevi, Greece}
\author{A.~Farricker}
\affiliation{Cockroft Institute (A36), Liverpool University, Warrington WA4 4AD, UK}
\author{E.~Fasoula}
\affiliation{Department of Physics, Aristotle University of Thessaloniki, Thessaloniki, Greece}
\affiliation{Center for Interdisciplinary Research and Innovation (CIRI-AUTH), Thessaloniki, Greece}
\author{T.~Fukuda}
\affiliation{Institute for Advanced Research, Nagoya University, Nagoya 464–8601, Japan}
\author{N.~Gazis}
\affiliation{European Spallation Source, Box 176, SE-221 00 Lund, Sweden}
\author{Th.~Geralis}
\affiliation{Institute of Nuclear and Particle Physics, NCSR Demokritos, Neapoleos 27, 15341 Agia Paraskevi, Greece}
\author{M.~Ghosh}
\affiliation{Center of Excellence for Advanced Materials and Sensing Devices, Ruđer Bo\v{s}kovi\'c Institute, 10000 Zagreb, Croatia}
\author{A.~Giarnetti}
\affiliation{Dipartimento di Matematica e Fisica, Universit\'a di Roma Tre, Via della Vasca Navale 84, 00146 Rome, Italy}
\author{G.~Gokbulut}
\affiliation{University of Cukurova, Faculty of Science and Letters, Department of Physics, 01330 Adana, Turkey}
\affiliation{Department of Physics and Astronomy, Ghent University, Proeftuinstraat 86, B-9000 Ghent, Belgium}
\author{C.~Hagner}
\affiliation{Institute for Experimental Physics, Hamburg University, 22761 Hamburg, Germany}
\author{L.~Hali\'c}
\affiliation{Center of Excellence for Advanced Materials and Sensing Devices, Ruđer Bo\v{s}kovi\'c Institute, 10000 Zagreb, Croatia}
\author{M.~Hooft}
\affiliation{Department of Physics and Astronomy, Ghent University, Proeftuinstraat 86, B-9000 Ghent, Belgium}
\author{K.~E.~Iversen}
\affiliation{Department of Physics, Lund University, P.O Box 118, 221 00 Lund, Sweden}
\author{N.~Jachowicz}
\affiliation{Department of Physics and Astronomy, Ghent University, Proeftuinstraat 86, B-9000 Ghent, Belgium}
\author{M.~Jensen}
\affiliation{European Spallation Source, Box 176, SE-221 00 Lund, Sweden}
\author{R.~Johansson}
\affiliation{European Spallation Source, Box 176, SE-221 00 Lund, Sweden}
\author{E.~Kasimi}
\affiliation{Department of Physics, Aristotle University of Thessaloniki, Thessaloniki, Greece}
\affiliation{Center for Interdisciplinary Research and Innovation (CIRI-AUTH), Thessaloniki, Greece}
\author{A.~Kayis Topaksu}
\affiliation{University of Cukurova, Faculty of Science and Letters, Department of Physics, 01330 Adana, Turkey}
\author{B.~Kildetoft}
\affiliation{European Spallation Source, Box 176, SE-221 00 Lund, Sweden}
\author{K.~Kordas}
\affiliation{Department of Physics, Aristotle University of Thessaloniki, Thessaloniki, Greece}
\affiliation{Center for Interdisciplinary Research and Innovation (CIRI-AUTH), Thessaloniki, Greece}
\author{B.~Kova\u{c}}
\affiliation{Center of Excellence for Advanced Materials and Sensing Devices, Ruđer Bo\v{s}kovi\'c Institute, 10000 Zagreb, Croatia}
\author{A.~Leisos}
\affiliation{Physics Laboratory, School of Science and Technology, Hellenic Open University, 26335, Patras, Greece }
\author{A.~Longhin}
\affiliation{Department of Physics and Astronomy "G. Galilei", University of Padova and INFN Sezione di Padova, Italy}
\author{C.~Maiano}
\affiliation{European Spallation Source, Box 176, SE-221 00 Lund, Sweden}
\author{S.~Marangoni}
\affiliation{University of Milano-Bicocca and INFN Sez. di Milano-Bicocca, 20126 Milano, Italy}
\author{J.~G.~Marcos}
\affiliation{Department of Physics and Astronomy, Ghent University, Proeftuinstraat 86, B-9000 Ghent, Belgium}
\author{C.~Marrelli}
\affiliation{CERN, 1211 Geneva 23, Switzerland}
\author{D.~Meloni}
\affiliation{Dipartimento di Matematica e Fisica, Universit\'a di Roma Tre, Via della Vasca Navale 84, 00146 Rome, Italy}
\author{M.~Mezzetto}
\affiliation{INFN Sez. di Padova, Padova, Italy}
\author{N.~Milas}
\affiliation{European Spallation Source, Box 176, SE-221 00 Lund, Sweden}
\author{R.~Moolya}
\affiliation{Institute for Experimental Physics, Hamburg University, 22761 Hamburg, Germany}
\author{J.L.~Mu\~noz}
\affiliation{Consorcio ESS-bilbao, Parque Cient\'{i}fico y Tecnol\'{o}gico de Bizkaia, Laida Bidea, Edificio 207-B, 48160 Derio, Bizkaia, Spain}
\author{K.~Niewczas}
\affiliation{Department of Physics and Astronomy, Ghent University, Proeftuinstraat 86, B-9000 Ghent, Belgium}
\author{M.~Oglakci}
\affiliation{University of Cukurova, Faculty of Science and Letters, Department of Physics, 01330 Adana, Turkey}
\author{T.~Ohlsson}
\email{Corresponding authors: S. Choubey (choubey@kth.se), T. Ohlsson (tohlsson@kth.se) and S. Vihonen (vihonen@kth.se)}
\affiliation{Department of Physics, School of Engineering Sciences, KTH Royal Institute of Technology, Roslagstullsbacken 21, 106 91 Stockholm, Sweden}
\affiliation{The Oskar Klein Centre, AlbaNova University Center, Roslagstullsbacken 21, 106 91 Stockholm, Sweden}
\author{M.~Olveg{\aa}rd}
\affiliation{Department of Physics and Astronomy, FREIA Division, Uppsala University, P.O. Box 516, 751 20 Uppsala, Sweden}
\author{M.~Pari}
\affiliation{Department of Physics and Astronomy "G. Galilei", University of Padova and INFN Sezione di Padova, Italy}
\author{D.~Patrzalek}
\affiliation{European Spallation Source, Box 176, SE-221 00 Lund, Sweden}
\author{G.~Petkov}
\affiliation{Sofia University St. Kliment Ohridski, Faculty of Physics, 1164 Sofia, Bulgaria}
\author{Ch.~Petridou}
\affiliation{Department of Physics, Aristotle University of Thessaloniki, Thessaloniki, Greece}
\affiliation{Center for Interdisciplinary Research and Innovation (CIRI-AUTH), Thessaloniki, Greece}
\author{P.~Poussot}
\affiliation{IPHC, Universit\'{e} de Strasbourg, CNRS/IN2P3, Strasbourg, France}
\author{A~Psallidas}
\affiliation{Institute of Nuclear and Particle Physics, NCSR Demokritos, Neapoleos 27, 15341 Agia Paraskevi, Greece}
\author{F.~Pupilli}
\affiliation{INFN Sez. di Padova, Padova, Italy}
\author{D.~Saiang}
\affiliation{Department of Civil, Environmental and Natural Resources Engineering Lule\aa~University~of~Technology, SE-971 87 Lulea, Sweden}
\author{D.~Sampsonidis}
\affiliation{Department of Physics, Aristotle University of Thessaloniki, Thessaloniki, Greece}
\affiliation{Center for Interdisciplinary Research and Innovation (CIRI-AUTH), Thessaloniki, Greece}
\author{A.~Scanu}
\affiliation{University of Milano-Bicocca and INFN Sez. di Milano-Bicocca, 20126 Milano, Italy}
\author{C.~Schwab}
\affiliation{IPHC, Universit\'{e} de Strasbourg, CNRS/IN2P3, Strasbourg, France}
\author{F.~Sordo}
\affiliation{Consorcio ESS-bilbao, Parque Cient\'{i}fico y Tecnol\'{o}gico de Bizkaia, Laida Bidea, Edificio 207-B, 48160 Derio, Bizkaia, Spain}
\author{G.~Stavropoulos}
\affiliation{Institute of Nuclear and Particle Physics, NCSR Demokritos, Neapoleos 27, 15341 Agia Paraskevi, Greece}
\author{R.~Tarkeshian}
\affiliation{European Spallation Source, Box 176, SE-221 00 Lund, Sweden}
\author{F.~Terranova}
\affiliation{University of Milano-Bicocca and INFN Sez. di Milano-Bicocca, 20126 Milano, Italy}
\author{T.~Tolba}
\affiliation{Institute for Experimental Physics, Hamburg University, 22761 Hamburg, Germany}
\author{E.~Trachanas}
\affiliation{European Spallation Source, Box 176, SE-221 00 Lund, Sweden}
\author{R.~Tsenov}
\affiliation{Sofia University St. Kliment Ohridski, Faculty of Physics, 1164 Sofia, Bulgaria}
\author{A.~Tsirigotis}
\affiliation{Physics Laboratory, School of Science and Technology, Hellenic Open University, 26335, Patras, Greece }
\author{S.~E.~Tzamarias}
\affiliation{Department of Physics, Aristotle University of Thessaloniki, Thessaloniki, Greece}
\affiliation{Center for Interdisciplinary Research and Innovation (CIRI-AUTH), Thessaloniki, Greece}
\author{M.~Vanderpoorten}
\affiliation{Department of Physics and Astronomy, Ghent University, Proeftuinstraat 86, B-9000 Ghent, Belgium}
\author{G.~Vankova-Kirilova}
\affiliation{Sofia University St. Kliment Ohridski, Faculty of Physics, 1164 Sofia, Bulgaria}
\author{N.~Vassilopoulos}
\affiliation{Institute of High Energy Physics (IHEP) Dongguan Campus, Chinese Academy of Sciences (CAS), Guangdong 523803, China}
\author{S.~Vihonen}
\email{Corresponding authors: S. Choubey (choubey@kth.se), T. Ohlsson (tohlsson@kth.se) and S. Vihonen (vihonen@kth.se)}
\affiliation{Department of Physics, School of Engineering Sciences, KTH Royal Institute of Technology, Roslagstullsbacken 21, 106 91 Stockholm, Sweden}
\affiliation{The Oskar Klein Centre, AlbaNova University Center, Roslagstullsbacken 21, 106 91 Stockholm, Sweden}
\author{J.~Wurtz}
\affiliation{IPHC, Universit\'{e} de Strasbourg, CNRS/IN2P3, Strasbourg, France}
\author{V.~Zeter}
\affiliation{IPHC, Universit\'{e} de Strasbourg, CNRS/IN2P3, Strasbourg, France}
\author{O.~Zormpa}
\affiliation{Institute of Nuclear and Particle Physics, NCSR Demokritos, Neapoleos 27, 15341 Agia Paraskevi, Greece}

\collaboration{ESSnuSB Collaboration}
\noaffiliation

\date{\today}

\begin{abstract}
\newpage
Atmospheric neutrinos provide a unique avenue to study neutrino interactions in matter. In this work, the prospects of constraining non-standard neutrino interactions with atmospheric neutrino oscillations are investigated for the proposed ESSnuSB far detector. By analyzing atmospheric neutrino samples equivalent to 5.4~Mt$\cdot$year exposure, it is found that ESSnuSB could be able to set the upper bounds $|\epsilon_{e\mu}^m| < 0.053, |\epsilon_{e\tau}^m| < 0.057, |\epsilon_{\mu\tau}^m| < 0.021, \epsilon_{ee}^m - \epsilon_{\mu\mu}^m < 0.075$ and $|\epsilon_{\tau\tau}^m - \epsilon_{\mu\mu}^m| < 0.031$ at $90\%$~CL, when the results are minimized for $\phi_{e\mu}^m, \phi_{e\tau}^m$ and $\phi_{\mu\tau}^m$ and normal ordering is assumed for neutrino masses. It is also shown that the presence of non-standard interactions could affect the sensitivities to neutrino mass ordering and $\theta_{23}^{}$ octant in comparison to the standard interaction scheme. The results of this work highlight the complementarity between atmospheric and accelerator neutrino programs in ESSnuSB.
\end{abstract}

\maketitle

\section{\label{sec:intro}Introduction}

The standard paradigm of neutrino physics states that the mixing between the neutrino flavor states $\nu_e, \nu_\mu, \nu_\tau$ and the neutrino mass states $\nu_1, \nu_2, \nu_3$ can be described with three mixing angles $\theta_{12}^{}$, $\theta_{13}^{}$, $\theta_{23}^{}$ and one charge-parity (CP) phase $\delta_{\rm CP}^{}$. The mixing among neutrino states gives rise to neutrino oscillations, with oscillation frequencies defined by the mass-squared differences $\Delta m_{21}^2 \equiv m_2^2 - m_1^2$ and $\Delta m_{31}^2 \equiv m_3^2 - m_1^2$, where $m_1, m_2, m_3$ are the definite neutrino masses. Extensive data from neutrino oscillation experiments with accelerator, reactor, solar and atmospheric neutrino sources have constrained the values of the mixing angles and mass-squared differences to 2--4\% precision at $1\sigma$ confidence level (CL)~\cite{Esteban:2024eli}, leaving only the sign of $\Delta m_{31}^2$, the octant of $\theta_{23}^{}$ and the value of $\delta_{\rm CP}^{}$ to be determined in future experiments. With the accumulation of experimental data, it is also becoming viable to analyze neutrino oscillations to search physics beyond the Standard Model (SM). One example of such tests is the search for non-standard interactions (NSI), in which case the SM interactions of neutrinos are complemented by additional charged-current (CC) or neutral-current (NC) interactions~\cite{Ohlsson:2012kf,Farzan:2017xzy}, resulting in non-standard effects in neutrino production, neutrino detection and neutrino propagation. It is also possible to realize NSI in other Lorentz structures, such as the ones emerging from scalar-mediated interactions~\cite{Ge:2018uhz}. As neutrino physics is about to enter an era of precision measurements, it is well-motivated to investigate the potential to find non-standard physics.

The European Spallation Source neutrino SuperBeam (ESSnuSB)~\cite{Alekou:2022emd} project investigates the feasibility to measure $\delta_{\rm CP}^{}$ near the second oscillation maximum. This can be achieved by generating high-power beams of muon neutrinos and muon antineutrinos with 5~MW beam power at the European Spallation Source (ESS) accelerator facility, which is currently under construction near Lund in Sweden. The long-baseline neutrino program in ESSnuSB is expected to measure $\delta_{\rm CP}^{}$ by at least 9$^\circ$ resolution at $1\sigma$~CL regardless of the true value of $\delta_{\rm CP}^{}$~\cite{Abele:2022iml,ESSnuSB:2023ogw}. The far detectors consist of two cylindrical Water Cherenkov neutrino detectors, which would be placed inside the mine in Zinkgruvan about 360~km from Lund. The combined fiducial mass of the two Water Cherenkov detectors is proposed to be 540~kt. Thanks to the 1~km rock overburden and the geographical location of Zinkgruvan, the Water Cherenkov detectors planned for the ESSnuSB far detector complex also have physics potential for various non-beam applications, including atmospheric neutrinos, geoneutrinos, solar neutrinos, supernova neutrinos and proton decay~\cite{ESSnuSB:2013dql}.

The prospects of observing signatures of CC-NSI and NC-NSI have previously been studied for ESSnuSB in its neutrino beam program, see \textit{e.g.} Refs.~\cite{Blennow:2015nxa,Delgadillo:2023lyp}. Scalar NSI has been investigated for ESSnuSB in Ref.~\cite{ESSnuSB:2023lbg}. There is also a possibility to examine NC-NSI effects through the observation of coherent-elastic neutrino-nucleus scattering (CE$\nu$NS), which could be studied by placing additional nuclear recoil detectors in the vicinity of ESS~\cite{Baxter:2019mcx,Chatterjee:2022mmu,Simon:2024xwb}. Furthermore, ESSnuSB has been shown to have sensitivity to other new physics scenarios, including neutrino oscillations with a light sterile neutrino~\cite{Ghosh:2019zvl}, invisible neutrino decay~\cite{Choubey:2020dhw}, CPT violation~\cite{Majhi:2021api}, non-unitary mixing~\cite{Chatterjee:2021xyu}, ultra-light scalar dark matter~\cite{Cordero:2022fwb}, quantum decoherence~\cite{Cheng:2022lys,ESSnuSB:2024yji} and long-range forces~\cite{ESSnuSB:2025shd}.

In the present work, we examine the opportunities to probe NC-NSI effects by observing atmospheric neutrinos in the far detector of ESSnuSB. The methodology will follow our previous work in Ref.~\cite{ESSnuSB:2024wet}. The atmospheric neutrino interactions in the ESSnuSB far detectors are simulated by generating a large set of Monte Carlo (MC) events with the \texttt{GENIE} neutrino event generator~\cite{Andreopoulos:2009rq,GENIE:2021npt} with the use of \texttt{ROOT}~\cite{Brun:1997pa}. The MC samples are analyzed with a Python-based analysis software, which incorporates neutrino oscillations with a reweighting algorithm. The analysis software also approximates the detector effects for ESSnuSB. The neutrino oscillation probabilities used in this work are computed with \texttt{GLoBES}~\cite{Huber:2004ka,Huber:2007ji} with an add-on that includes NSI effects in probability calculation~\cite{Kopp:2008ds}. The sensitivities to constrain the NSI parameters relevant for matter effects are derived by analyzing the MC samples that are generated for atmospheric neutrinos and normalized to 5.4~Mt$\cdot$year exposure. Moreover, the implications on the determinations of the neutrino mass ordering and the octant of $\theta_{23}$ are also addressed in this work. 

The results presented in this work are complementary to the long-baseline neutrino oscillation program of the ESSnuSB project. The complementarity between the neutrino beam and atmospheric neutrinos was previously studied for MEMPHYS-like neutrino detectors in Ref.~\cite{Blennow:2019bvl}. In the present work, we examine the prospects of atmospheric neutrinos in the neutrino detectors that are described in the ESSnuSB Design Study~\cite{ESSnuSB:2023ogw}, while approximating the conditions inside the mine in Zinkgruvan.

This article is organized as follows. In Section~\ref{sec:nsi-theory}, the theoretical formalism for neutrino oscillations is reviewed for NSI. The effects of NC-NSI on the neutrino oscillation probabilities are discussed in this section with analytical formulas. In Section~\ref{sec:probabilities}, neutrino oscillation probabilities are investigated numerically and the effect of each NSI parameter is discussed for atmospheric neutrino oscillations. The experimental configuration and the numerical analysis of the MC samples are summarized for ESSnuSB in Section~\ref{sec:configuration}, whereas the main results of this work are shown in Sections~\ref{sec:nsi_constraints} and~\ref{sec:nsi_effects}. Finally, in Section~\ref{sec:concl}, we present our concluding remarks.

\section{\label{sec:nsi-theory}Non-Standard Interactions}

In this section, we present a brief review of the NSI formalism, focusing on the $V-A$ Lorentz structure. We also examine the effect of individual NSI parameters on the neutrino oscillation probabilities relevant for atmospheric neutrinos.

\subsection{\label{sec:nsi-theory:formalism}Review of the non-standard interaction formalism}

Non-standard neutrino interactions can be parameterized in the low-energy regime in terms of effective Lagrangians, which can be written separately for CC and NC interactions. The CC and NC Lagrangians responsible for NSI effects, hereafter denoted as CC-NSI and NC-NSI, respectively, are given as
\begin{align}
    \mathcal{L}^{\rm NSI}_{\rm CC} &= -2\sqrt{2} G_F \epsilon_{\alpha \beta}^{ff', C} (\bar{\nu}_\alpha^{} \gamma^\mu_{} P_L^{} \nu_\beta^{}) (\bar{f} \gamma^\mu P_C^{} f'),\label{eq:2.1}\\      
    \mathcal{L}^{\rm NSI}_{\rm NC} &= -2\sqrt{2} G_F \epsilon_{\alpha \beta}^{f, C} (\bar{\nu}_\alpha^{} \gamma^\mu_{} P_L^{} \nu_\beta^{}) (\bar{f} \gamma^\mu_{} P_C^{} f)\label{eq:2.2},
\end{align}
where $f, f^{'} = \ell_i^{}, u_i^{}, d_i^{}$ stand for charged leptons and quarks ($i = 1, 2, 3$), $\alpha, \beta = e, \mu, \tau$ refer to the flavors of the incoming and outgoing neutrinos, and $G_F^{}$ is the Fermi coupling constant. The chirality of the charged lepton interaction is determined by $C = L, R$, whereas $P_L^{} = (1-\gamma_5^{})/2$ and $P_R = (1+\gamma_5^{})/2$ are the projection operators for the left-handed and right-handed gauge interactions, respectively. The relative strengths of the NSI are parameterized by the dimensionless parameters $\epsilon_{\alpha \beta}^{ff', C}$ and $\epsilon_{\alpha \beta}^{f, C}$. 

The CC-NSI and NC-NSI Lagrangians presented in equations~(\ref{eq:2.1}) and (\ref{eq:2.2}) affect neutrino oscillations in different manners. The CC-NSI Lagrangian is relevant for the neutrino interactions that take place both in the source and the detector, whereas the NC-NSI Lagrangian affects neutrino interactions with ambient matter. 

In the presence of CC-NSI, the neutrino states that are produced and detected are not necessarily pure flavor states. Instead, the neutrino states involved in the production and detection processes become superpositions of the different neutrino flavor states. This is the case if the matrix that spans $\epsilon^{ff', C}_{}$ has non-diagonal elements. The incoherent production and detection of the neutrino flavor states due to the CC-NSI Lagrangian are often parameterized with the source NSI parameters $\epsilon_{\alpha \beta}^s$ and detection NSI parameters $\epsilon_{\alpha \beta}^d$, which can be complex for all neutrino flavors $\alpha, \beta = e, \mu, \tau$. The initial and final states of a neutrino that is detected in a neutrino experiment can therefore be written as
\begin{align}\label{eq:2.3}
|\nu_\alpha^s\rangle &= |\nu_\alpha^{}\rangle + \sum_{\gamma = e, \mu, \tau} \epsilon^s_{\alpha \gamma}|\nu_\gamma^{}\rangle,\\\label{eq:2.3b}
\langle\nu_{\beta}^d| &= \langle\nu_\beta^{}| + \sum_{\gamma = e, \mu, \tau} \epsilon^d_{\gamma \beta} \langle\nu_\gamma^{}|,
\end{align}
where $|\nu_\alpha^s\rangle$ and $\langle\nu_{\beta}^d|$ are the impure states that are respectively involved in the neutrino production and detection processes, in which $\alpha, \beta = e, \mu, \tau$ are the corresponding neutrino flavors. The exact dependence of the source and detector NSI parameters defined in equations~(\ref{eq:2.3}) and (\ref{eq:2.3b}) and those provided for the CC-NSI Lagrangian in equation~(\ref{eq:2.1}) depends on the processes involved in the neutrino production and detection.

The NC-NSI Lagrangian presented in equation~(\ref{eq:2.2}) is relevant for neutrino propagation. The NSI effects affecting neutrino propagation through matter are effectively parameterized by the so-called matter NSI parameters, which are given as a sum of the left-handed and right-handed NSI parameters in the NC-NSI Lagrangian:
\begin{equation}
    \label{eq:2.4}
    \epsilon_{\alpha \beta}^m = \sum_f^{} \left( \epsilon_{\alpha \beta}^{f, L} + \epsilon_{\alpha \beta}^{f, R} \right) \frac{N_f^{}}{N_e^{}} = \sum_f^{} \epsilon_{\alpha \beta}^f \frac{N_f^{}}{N_e^{}},
\end{equation}
where $N_f^{}$ stands for the number density of the fermion $f$ in matter through which the neutrino propagates. In a similar manner, $N_e^{}$ denotes the number density of electrons in matter. Since there are approximately two nucleons per electron in Earth matter, neutrinos propagating inside the Earth are sensitive to the combination 
\begin{equation}
    \label{eq:2.4b}
    \epsilon_{\alpha \beta}^m \simeq 3 \epsilon_{\alpha \beta}^u + 3 \epsilon_{\alpha \beta}^d + \epsilon_{\alpha \beta}^e,
\end{equation}
where $\epsilon_{\alpha \beta}^u, \epsilon_{\alpha \beta}^d$ and $\epsilon_{\alpha \beta}^e$ are the NC-NSI Lagrangian coefficients corresponding to up quarks, down quarks and electrons, respectively. The effective Hamiltonian responsible for neutrino propagation in matter therefore becomes
\begin{equation}
\label{eq:2.5}
H(E_\nu^{}) = \frac{1}{2E_{\nu}^{}}\left[U
\left(
\begin{array}{ccc}
0 \,\,\, & 0 & 0 \\
0 \,\,\, & \Delta m_{21}^2 & 0\\
0 \,\,\, & 0 & \Delta m_{31}^2 
\end{array}
\right) U^{\dagger}
+ A
\left(
\begin{array}{ccc}
1+\epsilon_{ee}^m & \epsilon_{e\mu}^m & \,\,\, \epsilon_{e\tau}^m \\
\epsilon_{e\mu}^{m*} & \epsilon_{\mu\mu}^m & \,\,\, \epsilon_{\mu\tau}^m \\
\epsilon_{e\tau}^{m*} & \epsilon_{\mu\tau}^{m*} & \,\,\, \epsilon_{\tau\tau}^m
\end{array}
\right)
\right],
\end{equation}
where $U$ is the well-known Pontecorvo-Maki-Nakagawa-Sakata matrix and $E_\nu^{}$ is the energy of the propagating neutrino. The standard matter effects give rise to the matter potential $A = 2 \sqrt{2} E_\nu^{} G_F^{} N_e^{}$, which is driven by the neutrino energy $E_\nu^{}$ and the electron number density $N_e^{}$. The corresponding Hamiltonian for the antineutrino oscillations can be obtained from equation~(\ref{eq:2.5}) with the transformations $\delta_{\rm CP}^{} \rightarrow - \delta_{\rm CP}^{}$ and $A \rightarrow - A$. The NSI effects related to neutrino propagation are parameterized with the six parameters $\epsilon_{ee}^m$, $\epsilon_{e\mu}^m$, $\epsilon_{e\tau}^m$, $\epsilon_{\mu\mu}^m$, $\epsilon_{\mu\tau}^m$ and $\epsilon_{\tau\tau}^m$. While the diagonal elements $\epsilon_{ee}^m$, $\epsilon_{\mu\mu}^m$ and $\epsilon_{\tau\tau}^m$ are real, the off-diagonal elements $\epsilon_{e\mu}^m$, $\epsilon_{e\tau}^m$ and $\epsilon_{\mu\tau}^m$ can even be complex. The probability for neutrino oscillations is given by $P_{\nu_\alpha^{} \rightarrow \nu_\beta^{}}(L, E_\nu) = | \langle \nu_\beta^d| e^{-i H(E_\nu^{}) L} | \nu_\alpha^s \rangle |^2$, where $L$ is the distance that neutrinos propagate.

In practice, neutrino oscillation experiments cannot probe the diagonal NSI parameters $\epsilon_{ee}^m, \epsilon_{\mu\mu}^m$ and $\epsilon_{\tau\tau}^m$ directly. Instead, experiments can test their differences, {\em e.g.} $\epsilon_{ee}^m - \epsilon_{\mu\mu}^m$ and $\epsilon_{\tau\tau}^m - \epsilon_{\mu\mu}^m$. This is owed to the fact that any real number can be subtracted from the diagonal elements of the Hamiltonian in equation~({\ref{eq:2.5}}) without affecting the neutrino oscillation probabilities. In addition, atmospheric neutrinos can be used to probe the off-diagonal NSI parameters $\epsilon_{e\mu}^m, \epsilon_{e\tau}^m$ and $\epsilon_{\mu\tau}^m$ in experiments.

\subsection{\label{sec:nsi-theory:implications}Effects on atmospheric neutrino oscillations}

The phenomenological consequences of the NSI parameters have been widely discussed in the literature, see {\em e.g.} Ref.~\cite{Ohlsson:2012kf,Farzan:2017xzy} for reviews. In the case of atmospheric neutrinos, the most relevant type of the three NSI discussed in this work is the matter NSI, which alter neutrino oscillation probabilities through neutrino propagation. In this subsection, we briefly discuss how the matter NSI parameters $\epsilon_{\alpha \beta}^m$ ($\alpha, \beta = e, \mu, \tau$) influence neutrino oscillation probabilities for atmospheric neutrinos.

The most relevant neutrino oscillation channels for atmospheric neutrinos are $\nu_e^{} \rightarrow \nu_\mu^{}, \nu_\mu^{} \rightarrow \nu_e^{}$ and $\nu_\mu^{} \rightarrow \nu_\mu^{}$. The corrections to neutrino oscillation probabilities in the $\nu_e^{} \rightarrow \nu_\mu^{}$ channel can be approximated with the analytical formula (derived from equation~(10) in Ref.~\cite{Kopp:2008ds}),
\begin{equation}
    \label{eq:prob_emu}
    \begin{split}
    \Delta P_{\nu_e^{} \rightarrow \nu_\mu^{}}^{\rm NSI} \simeq 
        &- 2 |\epsilon_{e \tau}^m| \sin 2\theta_{13}^{} \sin 2\theta_{23}^{} \sin \theta_{23}^{} \sin(\delta_{\rm CP}^{} + \phi_{e \tau}^m) \sin\Delta \sin(\hat{A}\Delta) \frac{\sin((1-\hat{A})\Delta)}{1-\hat{A}} \\
        &- 2 |\epsilon_{e \tau}^m| \sin 2\theta_{13}^{} \sin 2\theta_{23}^{} \sin \theta_{23}^{} \cos(\delta_{\rm CP}^{} + \phi_{e \tau}^m) \cos\Delta \cos(\hat{A}\Delta) \frac{\sin((1-\hat{A})\Delta)}{1-\hat{A}} \\
        &+ 2 |\epsilon_{e \tau}^m| \sin 2\theta_{13}^{} \sin 2\theta_{23}^{} \sin \theta_{23}^{} \cos(\delta_{\rm CP}^{} + \phi_{e \tau}^m) \hat{A} \frac{\sin^2_{}((1-\hat{A})\Delta)}{(1-\hat{A})^2},
    \end{split}    
\end{equation}
where $\Delta = \Delta m_{31}^2 L / (4 E_\nu^{})$ and $\hat{A} = 2\sqrt{2} E_\nu^{} G_F^{} N_e^{} / \Delta m_{31}^2$. Here, $\Delta P_{\nu_e \rightarrow \nu_\mu}^{\rm NSI} \equiv P_{\nu_e \rightarrow \nu_\mu}^{\rm NSI} - P_{\nu_e \rightarrow \nu_\mu}^{\rm SI}$ is the difference between the oscillation probabilities derived for the matter NSI parameter $\epsilon_{e\tau}^m \equiv |\epsilon_{e\tau}^m|\exp{(-i \phi_{e\tau}^m)}$, which is assumed to be a small parameter, and for standard interactions (SI). The neutrino oscillation probabilities are expressed up to first order in $\sin \theta_{13}^{}$ and $\epsilon_{e\tau}^m$ combined. An analogous expression can be derived for the $\nu_e^{} \rightarrow \nu_{\tau}^{}$ channel~\cite{Kopp:2008ds}.

The analytical formula in equation~(\ref{eq:prob_emu}) shows the dependence of $P_{\nu_e \rightarrow \nu_\mu}^{}$ on both $|\epsilon_{e\tau}^m|$ and $\phi_{e\tau}^m.$ Neutrino oscillation probabilities were also examined analytically for the matter NSI parameters in Ref.~\cite{Ribeiro:2007ud}, where analytical formulas were derived separately for $\epsilon_{e\mu}^m$ and $\epsilon_{e\tau}^m.$ In the case of $\epsilon_{e\mu}^m,$ the corrections to the standard probability $P_{\nu_e \rightarrow \nu_\mu}^{\rm SI}$ appear in the first-order expressions of $\sin \theta_{13}$ and $\epsilon_{e\mu}^m,$ which indicates that $\epsilon_{e\mu}^m$ contributes to $P_{\nu_e \rightarrow \nu_\mu}^{}$ with similar strength as $\epsilon_{e\tau}^m.$

For the muon neutrino survival probability $P_{\nu_\mu^{} \rightarrow \nu_\mu^{}}$, the contributions from the matter NSI parameters arise mainly from the off-diagonal parameter $\epsilon_{\mu\tau}^m$. The approximate probability is given by~\cite{Choubey:2014iia}
\begin{equation}
    \label{eq:2.3z}
    \begin{split}
    \Delta P_{\nu_\mu \rightarrow \nu_\mu}^{\rm NSI}
    \simeq
    &-2|\epsilon_{\mu \tau}^m|\cos\phi_{\mu\tau}^m \hat{A} \sin 2\theta_{23}^{} \left(\sin^2_{} 2\theta_{23}^{} \Delta \sin2\Delta + 2\cos^2_{} 2\theta_{23}^{} \sin^2_{} \Delta \right)\\
    &+\left(|\epsilon_{\mu\mu}^m|-|\epsilon_{\tau\tau}^m|\right)\hat{A}\sin^2_{} 2\theta_{23}^{}\cos2\theta_{13}^{}\left(\Delta\sin2\Delta-2\sin^2\Delta\right).
    \end{split}
\end{equation}
It can be noted from equation~(\ref{eq:2.3z}) that the muon neutrino survival probability is primarily affected by the matter NSI parameter $\epsilon_{\mu \tau}^m$. The second most important quantity is the difference between the absolute values of the diagonal NSI parameters $\epsilon_{\mu\mu}^m$ and $\epsilon_{\tau\tau}^m$, although their relative impact is not as strong as that of $\epsilon_{\mu\tau}^m$ due to the negative sign in the $-2\sin^2\Delta$ term. The relatively large effect on the muon neutrino survival probability due to $\mu-\tau$ sector parameters such as $\epsilon_{\mu\tau}^m$, $\epsilon_{\tau\tau}^m$ and $\epsilon_{\mu\mu}^m$ comes mainly from the enhanced oscillation in the tau neutrino appearance channel $\nu_\mu \rightarrow \nu_\tau$ at higher neutrino energies.

Matter NSI parameters have been studied in a number of atmospheric neutrino oscillation experiments. For the off-diagonal NSI parameters $\epsilon_{e\mu}^m$ and $\epsilon_{e\tau}^m$, the most stringent constraints to date have been obtained by ORCA6~\cite{KM3NeT:2024pte} that sets the upper bounds $|\epsilon_{e\mu}^m| < 0.056$ and $|\epsilon_{e\tau}^m| < 0.074$, which are reported at 90\%~CL and assume real-valued matter NSI parameters that affect only down quarks. The constraints obtained for complex-valued NSI parameters are found to be of similar strength~\cite{KM3NeT:2024pte}. Similarly, the last off-diagonal NSI parameter $\epsilon_{\mu\tau}^m$ has been constrained by IceCube~\cite{IceCube:2022ubv}, where high-energy neutrino data indicate that $|\epsilon_{\mu\tau}^m| < 0.0041$. Finally, the effective NSI parameter $\epsilon_{\tau\tau}^m - \epsilon_{\mu\mu}^m$ has been probed most precisely by the DeepCore data~\cite{IceCubeCollaboration:2021euf} that set the upper bound at $|\epsilon_{\tau\tau}^m - \epsilon_{\mu\mu}^m| < 0.014$. Other noteworthy upper bounds on matter NSI parameters for atmospheric neutrinos have also been reported by ANTARES and Super-Kamiokande collaborations~\cite{ANTARES:2021crm,Super-Kamiokande:2011dam}. Global constraints from neutrino oscillation and scattering experiments have been reviewed in Ref.~\cite{Coloma:2023ixt}.

\section{\label{sec:probabilities}Oscillograms with non-standard interactions}

In this section, the neutrino oscillation probabilities are reviewed in presence of NSI. For concreteness, we focus on the most relevant NSI parameters for atmospheric neutrinos, which are the matter NSI parameters $\epsilon_{\mu\tau}^m, \epsilon_{e\tau}^m, \epsilon_{e\mu}^m, \epsilon_{\mu\mu}^m$ and $\epsilon_{\tau\tau}^m$.

In this work, the neutrino oscillation probabilities are computed adopting the widely-used \texttt{GLoBES} software~\cite{Huber:2004ka,Huber:2007ji}. The probabilities are calculated using the \texttt{snu} add-on~\cite{Kopp:2008ds}, which extends the framework of \texttt{GLoBES} to NSI. We apply this framework to study the effect of matter NSI parameters in three-flavor neutrino oscillations. The matter effects are taken into account with a PREM profile~\cite{Dziewonski:1981xy} of 81 steps, which is sufficient to approximate the matter density profile through the Earth. The baseline lengths $L$ are computed for the neutrino cosine zenith angles $\cos \theta_z^{}$ with the well-known formula,
\begin{equation}
    \label{eq:3.0}
    L = \sqrt{(R + h)^2 + (R - d)^2 \sin^2 \theta_z^{}} - (R - d) \cos \theta_z^{},
\end{equation}
where $R$ is the Earth radius, $h$ is the neutrino production height and $d$ is the depth at which the neutrino detector is located. To approximate the conditions for the mine in Zinkgruvan, we set the depth parameter $d = 1$~km. For the remaining parameters, we assume $R = 6371$~km and $h = 15$~km. For the neutrino oscillation parameters, we follow the global best-fit values from \texttt{NuFit 6.0}~\cite{Esteban:2024eli,NuFIT:6.0}. The current best-fit values and the corresponding uncertainties are summarized in Table~\ref{tab:bestfits}.
\begin{table}[!t]
\begin{center}
\begin{tabular}{|c|c|c|c|}\hline
{\bf Parameter} & {\bf Central value} & {\bf Uncertainty (1\,$\sigma$~CL)} & {\bf Allowed range (3\,$\sigma$~CL)} \\ \hline
\rule{0pt}{3ex}$\sin^2 \theta_{12}^{}$ & $0.308$ & $3.8\%$ & $0.275 \rightarrow 0.345$ \\ 
\rule{0pt}{3ex}$\sin^2 \theta_{13}^{}$ & $0.02215$ & $2.7\%$ & $0.02030 \rightarrow 0.02388$ \\ 
\rule{0pt}{3ex}$\sin^2 \theta_{23}^{}$ & $0.470$ & $5.3\%$ & $0.435 \rightarrow 0.585$ \\ 
\rule{0pt}{3ex}$\delta_\text{CP}$ [$^\circ$] & $212$ & $18.9\%$ & $124 \rightarrow 364$ \\ 
\rule{0pt}{3ex}$\Delta m_{21}^2$ [10$^{-5}$ eV$^2$] & $7.49$ & $2.5\%$ & $6.92 \rightarrow 8.05$ \\ 
\rule{0pt}{3ex}$\Delta m_{31}^2$ [10$^{-3}$ eV$^2$] & $2.513$ & $0.8\%$ & $2.451 \rightarrow 2.578$ \\ \hline
\end{tabular}
\end{center}
\caption{\label{tab:bestfits} The best-fit values of the neutrino oscillation parameters as determined by neutrino experiments~\cite{Esteban:2024eli,NuFIT:6.0}. The values and their relative uncertainties are presented at $1\sigma$~CL assuming normal mass ordering. The allowed values are also shown for each parameter at $3\sigma$~CL. The values are based on \texttt{NuFit~6.0} best-fit results~\cite{NuFIT:6.0} including Super-Kamiokande data.}
\end{table}

\begin{figure}[!t]
    \centering
    \includegraphics[width=1.0\linewidth]{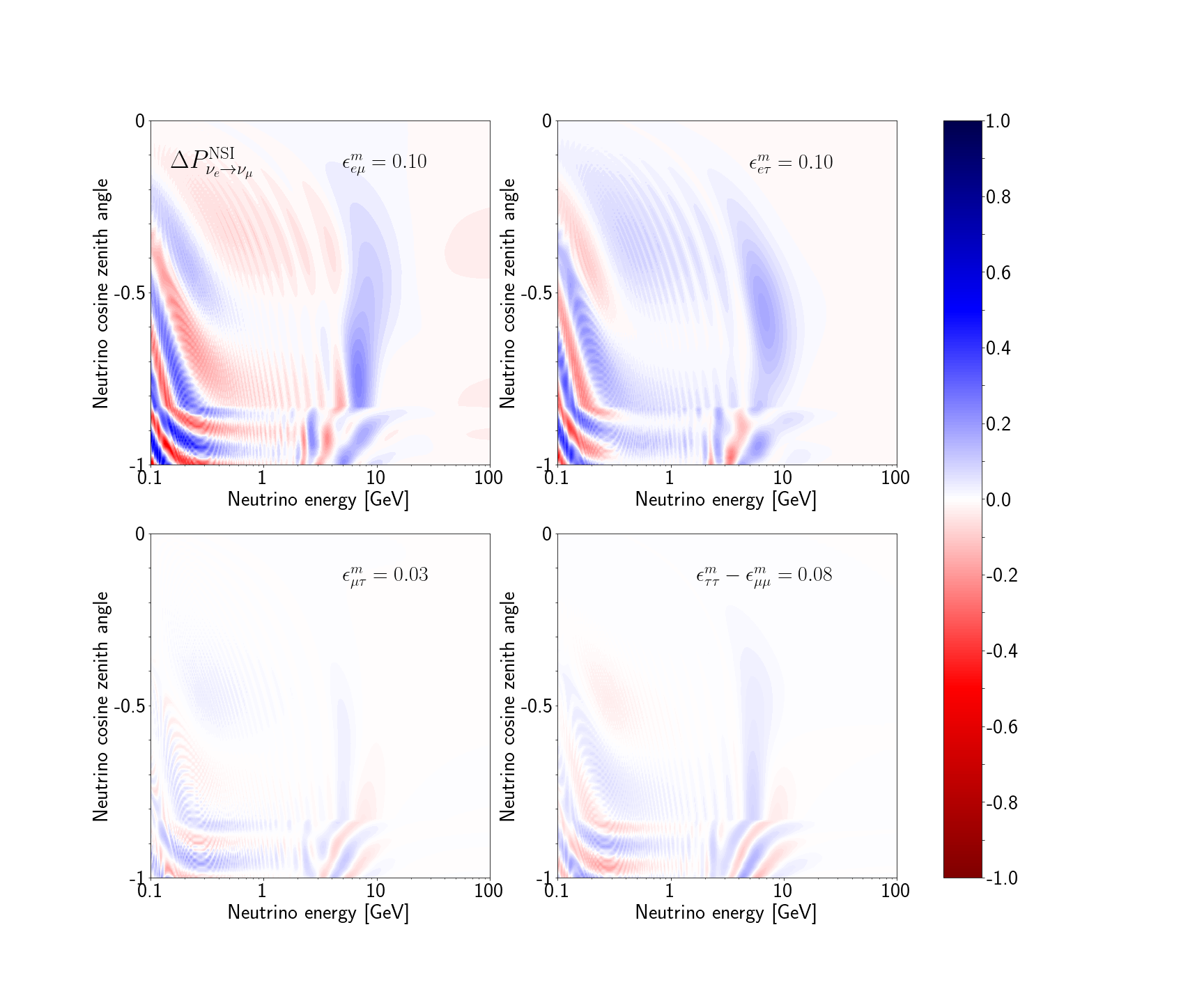}
    \vspace{-1.5cm}
    \caption{Probability differences for the neutrino oscillation probability $P_{e\mu}^{}$ owing to NSI effects in neutrino propagation. The probability difference is given by $\Delta P_{\nu_e \rightarrow \nu_\mu}^{\rm NSI} = P_{\nu_e \rightarrow \nu_\mu}^{\rm NSI} - P_{\nu_e \rightarrow \nu_\mu}^{\rm SI}$. In each panel, one NSI parameter is assigned a non-zero value for $P_{\nu_e \rightarrow \nu_\mu}^{\rm NSI}$, while the remaining NSI parameters are kept at zero. For $P_{\nu_e \rightarrow \nu_\mu}^{\rm SI}$, all NSI parameters are zero. The color bar indicates probability differences between $-100\%$ and $100\%$. The probability differences are computed assuming normal ordering for the neutrino masses.}
    \label{fig:nsi_ogram_Pemu}
\end{figure}

The effect of the matter NSI parameters $\epsilon_{e\mu}^m, \epsilon_{e\tau}^m, \epsilon_{\mu\tau}^m$ and $\epsilon_{\tau\tau}^m-\epsilon_{\mu\mu}^m$ on the neutrino oscillation probability $P_{\nu_e \rightarrow \nu_\mu}^{}$ is illustrated in Figure~\ref{fig:nsi_ogram_Pemu}. The colors show the probability differences that are computed for the probabilities with one non-zero NSI parameter, $P_{\nu_e \rightarrow \nu_\mu}^{\rm NSI}$, and with no NSI effects, $P_{\nu_e \rightarrow \nu_\mu}^{\rm SI}$. The probability difference $\Delta P_{\nu_e^{} \rightarrow \nu_\mu^{}}^{\rm NSI} \equiv P_{\nu_e^{} \rightarrow \nu_\mu^{}}^{\rm NSI} - P_{\nu_e^{} \rightarrow \nu_\mu^{}}^{\rm SI}$ is shown as a function of the neutrino energy $E_\nu^{}$ and the neutrino cosine zenith angle $\cos \theta_z^{}$. The color bar encodes probability differences between $-100\%$ and $100\%$. Without loss of generality, we have set the complex phases $\phi_{e\mu}^m,\phi_{e\tau}^m$ and $\phi_{\mu\tau}^m$ to zero. The magnitudes of the effective matter NSI parameters, {\em i.e.}, $|\epsilon_{e\mu}^m|, |\epsilon_{e\tau}^m|, |\epsilon_{\mu\tau}^m|, |\epsilon_{ee}^m - \epsilon_{\mu\mu}^m|$ and $|\epsilon_{\tau\tau}^m - \epsilon_{\mu\mu}^m|$, have been set to the values indicated in the panels, whereas the other NSI parameters are set to zero. The values chosen for the matter NSI parameters are within the $99\%$~CL limits of the global fit analysis~\cite{Coloma:2023ixt}. Figure~\ref{fig:nsi_ogram_Pemu} shows that the transition probability $P_{\nu_e \rightarrow \nu_\mu}^{}$ is enhanced by the NSI effects in multiple parts of the neutrino energy spectrum. For the off-diagonal NSI parameters $\epsilon_{e\mu}^m$ and $\epsilon_{e\tau}^m$, the probability is affected across the neutrino energy spectrum, but the most significant changes are found for sub-GeV neutrino energies $E_\nu < 1$~GeV. However, the neutrino energy and the neutrino cosine zenith angle become challenging to reconstruct in the sub-GeV region, and greater significance is therefore expected from multi-GeV neutrino energies. On the other hand, the matter NSI parameters $\epsilon_{\mu\tau}^m$ and $\epsilon_{\tau\tau}^m - \epsilon_{\mu\mu}^m$ are shown to have smaller effects on the transition probability than $\epsilon_{e\mu}^m$ and $\epsilon_{e\tau}^m$. For $\epsilon_{\mu\tau}^m$ and $\epsilon_{\tau\tau}^m - \epsilon_{\mu\mu}^m$, the inclusion of the matter NSI parameters results in percent-level corrections for neutrino energies $E_\nu^{} \lesssim 10$~GeV. The enhancements are the strongest near the direction where neutrinos propagate through the center of the Earth. However, the experimental constraints from IceCube and ORCA6~\cite{IceCube:2022ubv,KM3NeT:2024pte} indicate that the upper bound on $|\epsilon_{\mu\tau}^m|$ is much smaller than the one shown in Figure~\ref{fig:nsi_ogram_Pemu}. When this upper bound is imposed on $\epsilon_{\mu\tau}^m,$ the difference $\Delta P_{\nu_e \rightarrow \nu_\mu}^{\rm NSI}$ is reduced to sub-percent level. It can therefore be concluded that $\epsilon_{\mu\tau}^m$ and $\epsilon_{\tau\tau}^m - \epsilon_{\mu\mu}^m$ do not contribute significantly to $\nu_e \rightarrow \nu_\mu$ oscillations, while $\epsilon_{e\mu}^m$ and $\epsilon_{e\tau}^m$ may lead to notable effects.
\begin{figure}[!t]
    \centering
    \includegraphics[width=1.0\linewidth]{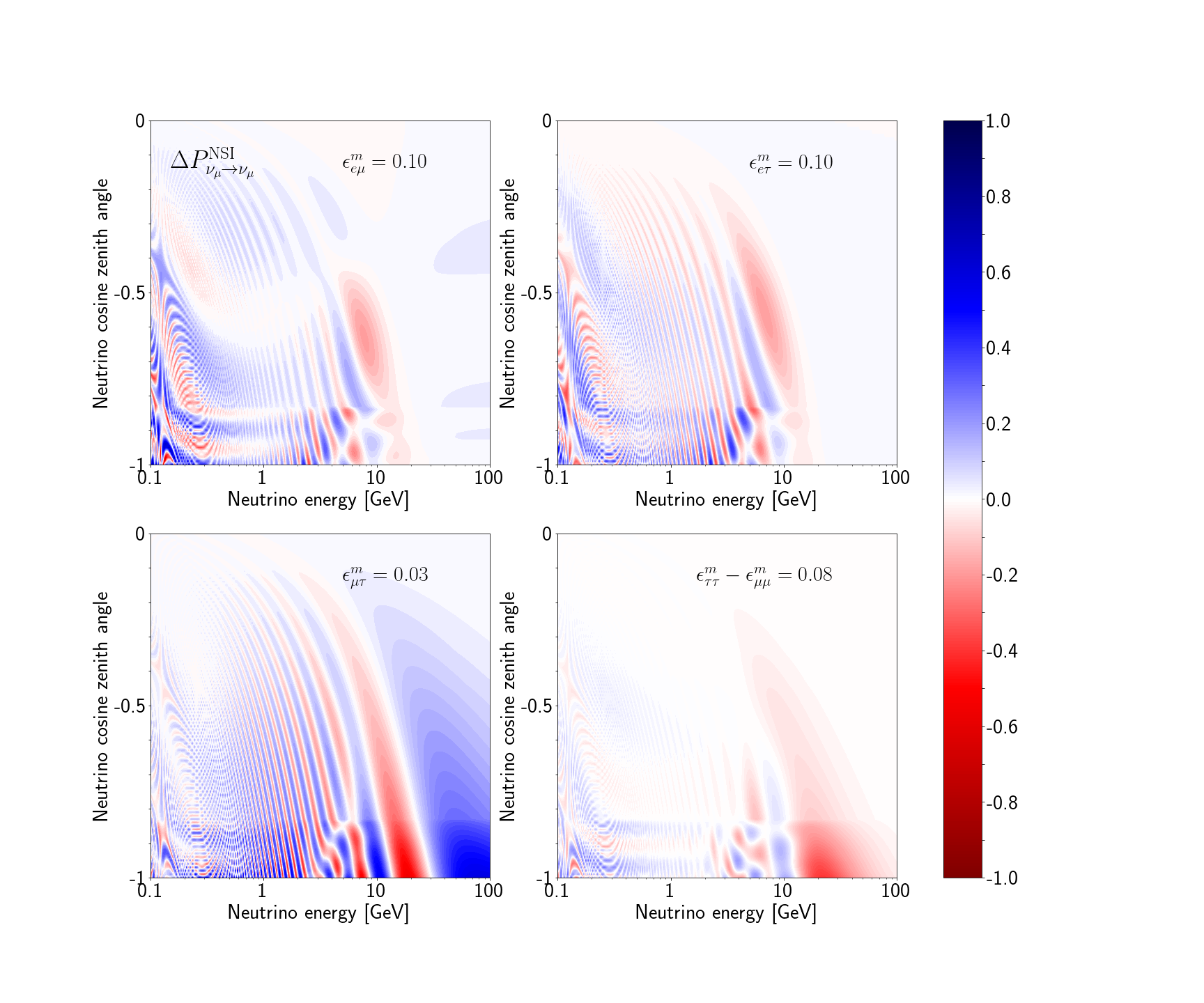}
    \vspace{-1.5cm}
    \caption{Probability difference $\Delta P_{\nu_\mu \rightarrow \nu_\mu}^{\rm NSI} = P_{\nu_\mu \rightarrow \nu_\mu}^{\rm NSI} - P_{\nu_\mu \rightarrow \nu_\mu}^{\rm SI}$ for the survival probability $P_{\nu_\mu^{} \rightarrow \nu_\mu^{}}$ due to the NSI effects in neutrino propagation. The probability $P_{\nu_\mu^{} \rightarrow \nu_\mu^{}}^{\rm NSI}$ is computed for one non-zero matter NSI parameter, whereas $P_{\nu_\mu^{} \rightarrow \nu_\mu^{}}^{\rm SI}$ corresponds to standard interactions. The color bar indicates probability differences between $-100\%$ and $100\%$. The probabilities are computed assuming normal ordering for the neutrino masses.}
    \label{fig:nsi_ogram_Pmumu}
\end{figure}

The $\mu-\tau$ sector is known to have a strong impact on the $\nu_\mu^{} \rightarrow \nu_\mu^{}$ oscillations. This was already pointed out in Section~\ref{sec:nsi-theory:implications}, where the analytical formula for $\Delta P_{\nu_\mu \rightarrow \nu_\mu}^{\rm NSI}$ in equation~(\ref{eq:2.3z}) has $|\epsilon_{\mu\tau}^m| \cos \phi_{\mu\tau}^m,|\epsilon_{\mu\mu}^m|$ and $|\epsilon_{\tau\tau}^m|$ in the leading term. In Figure~\ref{fig:nsi_ogram_Pmumu}, the difference between the probabilities $P_{\nu_\mu \rightarrow \nu_\mu}^{\rm NSI}$ and $P_{\nu_\mu \rightarrow \nu_\mu}^{\rm SI}$ is shown for non-zero values of $\epsilon_{e\mu}^m, \epsilon_{e\tau}^m, \epsilon_{\mu\tau}^m$ and $\epsilon_{\tau\tau}^m - \epsilon_{\mu\mu}^m$. We have again set the complex phases $\phi_{e\mu}^m, \phi_{e\tau}^m$ and $\phi_{\mu\tau}^m$ to zero. This figure shows that $\epsilon_{e\mu}^m$ alters the survival probability across the neutrino energies, whereas $\epsilon_{e\tau}^m$ has the most notable effects for neutrino energies $E_\nu^{} \lesssim 10$~GeV. In contrast, $\epsilon_{\mu\tau}^m$ and $\epsilon_{\tau\tau}^m - \epsilon_{\mu\mu}^m$ have the most significant effects for much higher neutrino energies, {\em i.e.}, $E_\nu^{} \sim 10~{\rm GeV}$. This is particularly true for $\epsilon_{\mu\tau}^m$, which yields significant contributions to the probability near $100$~GeV and beyond. For $\epsilon_{\tau\tau}^m - \epsilon_{\mu\mu}^m$, the highest impact appears around $E_\nu^{} \sim 25$~GeV. In both cases, the observed changes in the probability arise from $\nu_\mu \rightarrow \nu_\tau$ oscillations, which are enhanced by strong matter effects in the Earth's core. Atmospheric neutrino oscillations are therefore very sensitive to matter NSI parameters involving both muon and tau flavors. This behavior is also expected from the analytical formula that is shown for $P_{\nu_\mu \rightarrow \nu_\mu}^{}$ in equation~(\ref{eq:2.3z}) in Section~\ref{sec:nsi-theory:implications}.

\section{\label{sec:configuration}Atmospheric neutrinos at the ESSnuSB far detectors}

The analysis of the atmospheric neutrino data is carried out with the MC samples generated for Ref.~\cite{ESSnuSB:2024wet}. The samples were generated by \texttt{GENIE}~\cite{Andreopoulos:2009rq,GENIE:2021npt} and consist of atmospheric neutrino interactions with the neutrino energies $E_\nu \in [0.1, 100]$~GeV assuming 100~years of data taking and the \texttt{ROOT} geometry of a Water Cherenkov detector equivalent with 540~kt fiducial mass volume. The atmospheric neutrino fluxes were adopted from Ref.~\cite{Honda:2015fha} for Pyh\"asalmi, averaging over the seasonal variation. The large size of the generated MC samples ensures that the fluctuations in the MC chain remain small, when the samples are re-normalized to the designated exposure in the ESSnuSB far detectors. In the current work, we assume the total exposure for ESSnuSB to be $5.4~{\rm Mt}\cdot{\rm years}$, which is equivalent to 10 years of data taking.

The statistical analysis of the simulated neutrino events is carried out with the following procedure. The MC samples generated by \texttt{GENIE} are first pre-processed to obtain neutrino events for the test and the true hypotheses. During this phase, neutrino oscillations are taken into account in the MC samples by using a dedicated reweighting algorithm. The reweighting algorithm classifies atmospheric neutrino events either into oscillated or unoscillated events by generating a random number $S \in [0, 1]$ and comparing it with appropriate neutrino oscillation probabilities. The neutrino oscillation probabilities are computed with \texttt{GLoBES}. More details on the calculation of the neutrino oscillation probabilities can be found in Section~\ref{sec:probabilities}. Once the MC samples are reweighted, the oscillated atmospheric neutrino events are organized into neutrino energy and neutrino cosine zenith angle bins. In the present work, we consider 100 equidistant bins for the neutrino energies $E_\nu \in [0, 100]$~GeV and 20 bins for the neutrino cosine zenith angles $\cos \theta_z^{} \in [-1, 1]$ with bin sizes $\Delta E_\nu = 1$~GeV and $\Delta \cos \theta_z^{} = 0.1$, respectively\footnote{In principle, the down-going atmospheric neutrino events receive background from cosmic muons. However, the contamination from cosmic muons is estimated to be small for the mine in Zinkgruvan~\cite{Alekou:2022emd}, where the rock overburden is 1~km.}. We also incorporate the detector efficiencies for each neutrino flavor. The detector efficiencies are adopted from Ref.~\cite{Alekou:2022emd}. On average, the detector efficiencies are $62\%$ for $\nu_\mu^{}$ events, $54\%$ for $\nu_e^{}$ events, $69\%$ for $\bar\nu_\mu^{}$ events and $68\%$ for $\bar\nu_e^{}$ events. The energy threshold is $0.1$~GeV in our numerical analysis. Finally, the binned MC events are smeared with Gaussian functions which are applied independently for neutrino energies $\Delta E_\nu$ and neutrino zenith angles $\cos \theta_z^{}$ with the smearing functions $\sigma_{E_\nu}$ and $\sigma_{\cos \theta_z^{}}$, respectively. For our baseline setup, we assume $30\%$ resolution for neutrino energies in the sub-GeV bins and $10\%$ resolution for neutrino energies in the multi-GeV bins. We also assume the neutrino zenith angle resolution to be $10^\circ$. As the conventional Water Cherenkov technology is not sensitive to the sign of the electric charge of the charged leptons, the MC events are separated into electron-like and muon-like events, where neutrino and antineutrino events corresponding to the same neutrino flavor are included in the same analysis bin. In other words, electron-like events contain both events involving $\nu_e$ and $\bar\nu_e$ interactions, whereas muon-like events contain events involving both $\nu_\mu$ and $\bar\nu_\mu$ interactions. The events include CC interactions as well as NC interactions. It is important to note that neutrino oscillations are only taken into account for events involving CC interactions, whereas events involving NC interactions are unoscillated events and therefore treated as unoscillated backgrounds. In principle, gadolinium-doping could be used in the ESSnuSB far detectors to acquire sensitivity to the sign of the electric charge for charged leptons. This option is currently being investigated in an ongoing study. In addition, it is assumed in our numerical analysis that the probability to mis-identify electron-like events as muon-like events and muon-like events as electron-like events is zero for the ESSnuSB far detectors, since its impact on the results is presumed to be small. We will also check this impact in the numerical analysis.

The analysis of binned atmospheric neutrino events is carried out with the likelihood function,
\begin{equation}
    \label{eq:3.1}
    \chi_{}^2 = 2 \sum_{n=1}^{2000} \left( E_n^{} - O_n^{} + O_n^{} \log \frac{O_n^{}}{E_n^{}} \right) + \sum_{i=1}^{5} \left( \frac{\zeta_i^{}}{\sigma_i^{}} \right)^2,
\end{equation}
where $E_n^{}$ and $O_n^{}$ correspond to the expected and observed neutrino events in the $n_{}^{\rm th}$ analysis bin, respectively. The electron-like samples and the muon-like samples are treated separately. The systematic uncertainties are taken into account with the well-known pull method~\cite{Fogli:2002pt}. In this work, we assume five different systematic uncertainties which are characterized by the pull-parameters $\zeta_i^{}$ and their standard deviations $\sigma_i^{}$, where $i = 1, 2,\ldots, 5.$ The systematic uncertainties are listed in Table~\ref{tab:systematics}. The test events in the analysis bin $n$ are computed as follows
\begin{equation}
    \label{eq:3.2}
    E_n^{} = E_{n, 0}^{} \left( 1 + \sum_{i=1}^{5} f_{n, i}^{} \zeta_i^{} \right),
\end{equation}
where $E_{n, 0}$ is the number of test events before the pull and $f_{n, i}$ are the coefficients that determine the weights of the pull parameter $\zeta_i$ in the analysis bin $n$. In our numerical analysis, $\chi_{}^2$ is computed separately for both electron-like and muon-like samples with equation~(\ref{eq:3.1}). The full $\chi_{}^2$ function is then obtained as the sum of the electron-like and muon-like components. Each sample is assumed to be pure, meaning that electron-like events consist purely of events involving $\nu_e$ and $\bar\nu_e$ interactions and muon-like events of $\nu_\mu$ and $\bar\nu_\mu$ interactions, respectively.

The evaluation of the systematic uncertainties is based on the methodology that was introduced in Ref.~\cite{Gonzalez-Garcia:2004pka}. We follow the implementation that has been used in {\em e.g.} Ref.~\cite{Gandhi:2007td,Ghosh:2012px,Choubey:2015xha}. There are five different systematic uncertainties in this approach: (i) flux normalization error, (ii) cross-section normalization error, (iii) zenith angle dependence error, (iv) energy tilt error and (v) detector efficiency error. Each of the five errors has its own pull parameter. It should be noted that electron-like events and muon-like events are analyzed with separate sets of nuisance parameters. The flux normalization error and the cross-section normalization error account for the uncertainties in the overall normalizations of the atmospheric neutrino fluxes and the neutrino-nucleus cross-sections. We treat the flux and cross-section normalization errors with conservative $20\%$ and $10\%$ uncertainties, respectively. Meanwhile, the zenith angle dependence error parametrizes the uncertainty on the zenith angle dependence. We compute the weights for the zenith angle dependence error as $5\%$ of the average neutrino cosine zenith angle in each analysis bin. The energy tilt error introduces a tilt factor which parametrizes potential deviations from the power-law assumption for the atmospheric neutrino fluxes. Finally, we include an overall $5\%$ error for the detector efficiencies.

The weights for the energy tilt error are obtained with the following procedure. We first compute the tilted atmospheric neutrino flux $\Phi_{\delta}^{} (E_{\nu}^{})$ as~\cite{Gonzalez-Garcia:2004pka}
\begin{equation}
    \label{eq:tilt_spectrum}
        \Phi_{\delta}^{} (E_{\nu}^{}) = \Phi_{\nu, 0}^{} \left(\frac{E_{\nu}}{E_{0}}\right)_{}^{\delta} \simeq \Phi_{\nu, 0}^{} \left( 1 + \delta \log \frac{E_{\nu}^{}}{E_{0}^{}} \right),
\end{equation}
where $\delta$ is a tilt factor, $E_{0}^{}$ is a reference energy and $\Phi_{\nu, 0}^{}$ represents the unaltered atmospheric neutrino flux. The weights for the energy tilt error are acquired by generating MC events for both the original and tilted atmospheric neutrino fluxes and calculating their ratio. For this work, we obtain the tilted atmospheric neutrino fluxes with equation~(\ref{eq:tilt_spectrum}) by setting $\delta = 5\%$ and $E_{0}^{} = 2$~GeV. The same treatment was used in our previous study~\cite{ESSnuSB:2024wet}.
\begin{table}[!t]
\begin{center}
\begin{tabular}{|c|c|}\hline
{\bf Systematic error} & {\bf Uncertainty} \\ \hline
\rule{0pt}{3ex}Flux normalization & $20\%$ \\
\rule{0pt}{3ex}Cross-section normalization & $10\%$ \\
\rule{0pt}{3ex}Zenith angle dependence & varies \\
\rule{0pt}{3ex}Energy tilt & varies \\
\rule{0pt}{3ex}Detector efficiency & $5\%$ \\ \hline
\end{tabular}
\end{center}
\caption{\label{tab:systematics}List of systematic uncertainties used in this work. See the text for the implementation of the zenith angle dependence and energy tilt errors. More information on the methodology can be found in Ref.~\cite{Gonzalez-Garcia:2004pka,Gandhi:2007td,Ghosh:2012px,Choubey:2015xha}.}
\end{table}

The minimization of the $\chi^2_{}$ function is done for the standard neutrino oscillation parameters $\theta_{23}, \Delta m_{31}^2$ and $\delta_{\rm CP}^{}$, which are allowed to vary within their allowed $3\sigma$~CL ranges. The parameters $\theta_{12}, \theta_{13}$ and $\Delta m_{21}^2$ are kept fixed at the central values. The latter are strictly constrained by the solar and reactor neutrino experiments and are not important for our numerical analysis. The central values and their $3\sigma$~CL ranges are adopted from the global best-fit values~\cite{Esteban:2024eli}, which are also listed in Table~\ref{tab:bestfits} in Section~\ref{sec:probabilities}. For the evaluation of the NSI effects, the minimization of $\chi_{}^2$ also includes one matter NSI parameter. The true values for the standard neutrino oscillation parameters are adopted from the central values. For simplicity, we have set the true values of $\delta_{\rm CP}^{}$ and the complex NSI phases $\phi_{e\mu}^m, \phi_{e\tau}^m, \phi_{\mu\tau}^m$ to zero in our numerical analysis. Note that no priors are used in this work.

\section{\label{sec:nsi_constraints}Constraining matter NSI parameters at ESSnuSB}

Atmospheric neutrinos can provide sensitive probes to NSI effects in neutrino propagation. Previously the prospects for probing matter NSI parameters with atmospheric neutrinos have been investigated for Hyper-Kamiokande~\cite{Fukasawa:2015jaa,Fukasawa:2016nwn,Kelly:2017kch}, INO~\cite{Choubey:2015xha,Choubey:2016gps} and PINGU~\cite{Choubey:2014iia}. Sensitivities to matter NSI parameters through atmospheric neutrino detection can also be expected for the far detectors of the reactor neutrino experiment JUNO~\cite{JUNO:2015zny} and the long-baseline experiment DUNE~\cite{DUNE:2020ypp}. Furthermore, future atmospheric neutrino data from the neutrino telescopes IceCube-Gen2~\cite{IceCube-Gen2:2020qha}, KM3NeT-ORCA~\cite{KM3Net:2016zxf} and P-One~\cite{P-ONE:2020ljt} can be expected to probe the parameter space for the matter NSI parameters even further. The atmospheric neutrino data expected for the ESSnuSB far detector would provide complementary results on the search for the matter NSI parameters.

\begin{figure}[!t]
    \centering
    \includegraphics[width=0.9\linewidth]{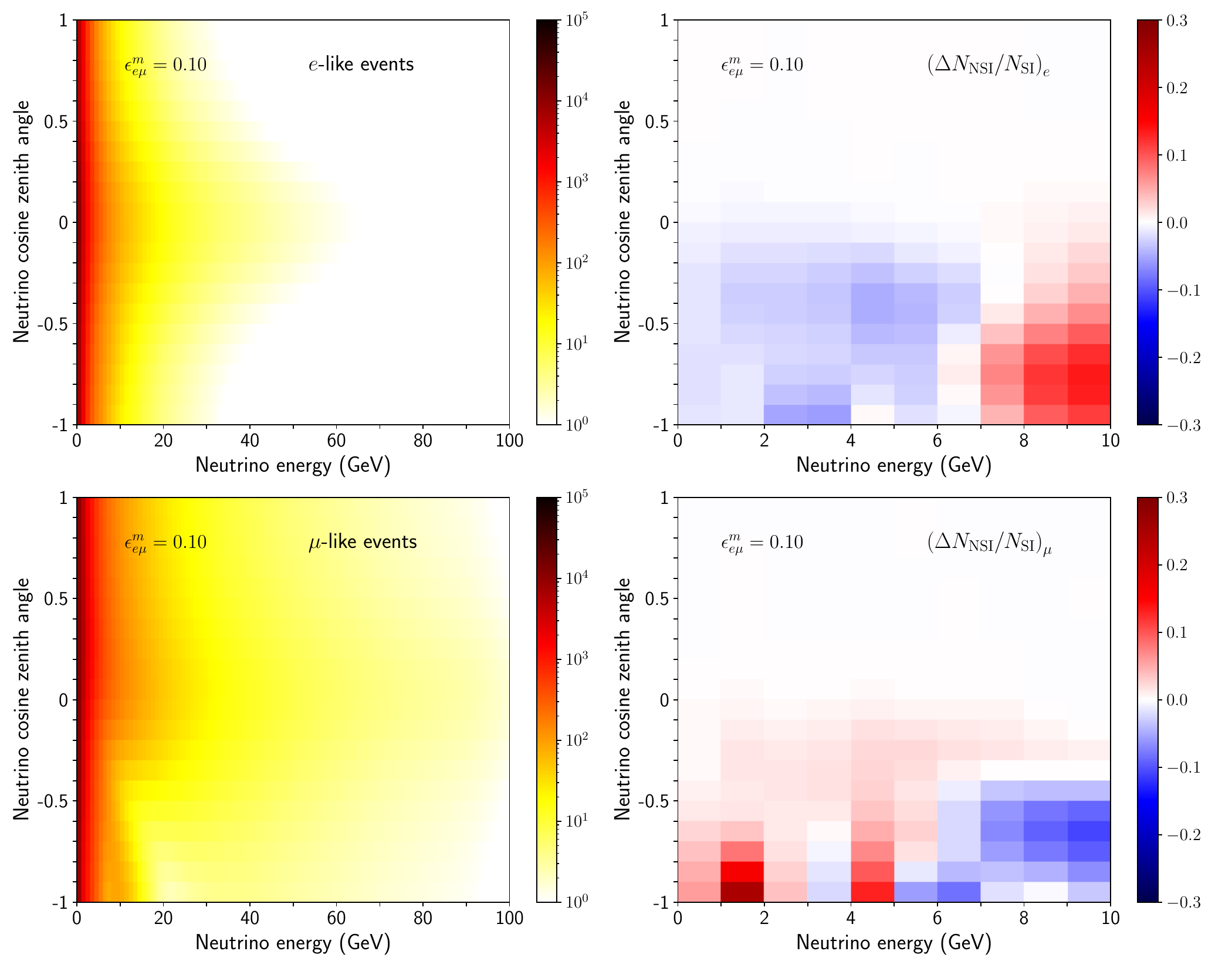}
    \caption{Expected number of atmospheric neutrino events in the ESSnuSB far detectors. Events are separated into $e$-like events (top-left) and $\mu$-like events (bottom-left) assuming $\epsilon_{e\mu}^m = 0.10.$ Also shown are the relative differences $\Delta N_{e}^{\rm NSI}/N_{e}^{\rm SI}$ for the $e$-like (top-right) and $\Delta N_{\mu}^{\rm NSI}/N_{\mu}^{\rm SI}$ for the $\mu$-like (bottom-right) samples, where $\Delta N_{e}^{\rm NSI}$ and $\Delta N_{\mu}^{\rm NSI}$ indicate the differences between events that are obtained with and without the NSI effects. Normal ordering is assumed for neutrino masses. Note that the relative differences are shown up to $10$~GeV, whereas the events are shown for the full neutrino energy range.}
    \label{fig:nsi_events_emu}
\end{figure}

The expected atmospheric neutrino spectrum for the ESSnuSB configuration is shown in Figure~\ref{fig:nsi_events_emu}. The top-left and bottom-left panels illustrate the expected number of events for the electron-like and muon-like samples, respectively. The events are computed for the NSI parameter value of $\epsilon_{e\mu}^m = 0.10$. The top-right and bottom-right panels show the corresponding relative differences $\Delta N_{e}^{\rm NSI}/N_{e}^{\rm SI} \equiv (N_{e}^{\rm NSI} - N_{e}^{\rm SI})/N_{e}^{\rm SI}$ and $\Delta N_{\mu}^{\rm NSI}/N_{\mu}^{\rm SI} \equiv (N_{\mu}^{\rm NSI} - N_{\mu}^{\rm SI})/N_{\mu}^{\rm SI}$ that have been obtained for the electron-like events and the muon-like events, respectively. The detector effects have been taken into account. The top-right and bottom-right panels show that the positive value of the NSI parameter $\epsilon_{e\mu}^m$ leads to notable changes in the expected electron-like and muon-like events. The effect is more pronounced for the multi-GeV bins, where the first oscillation maximum also occurs for atmospheric neutrinos. The aforementioned changes indicate that the atmospheric neutrinos that are expected for the ESSnuSB far detectors would have a moderate sensitivity to the off-diagonal NSI parameter $\epsilon_{e\mu}^m$. For the remaining matter NSI parameters, the relative event differences are presented in Appendix~\ref{Appendix:Events}. The effect of the zenith angle dependence error on the expected electron-like and muon-like samples is also discussed in the same appendix.

\begin{figure}[!t]
    \centering
    \includegraphics[width=1.0\linewidth]{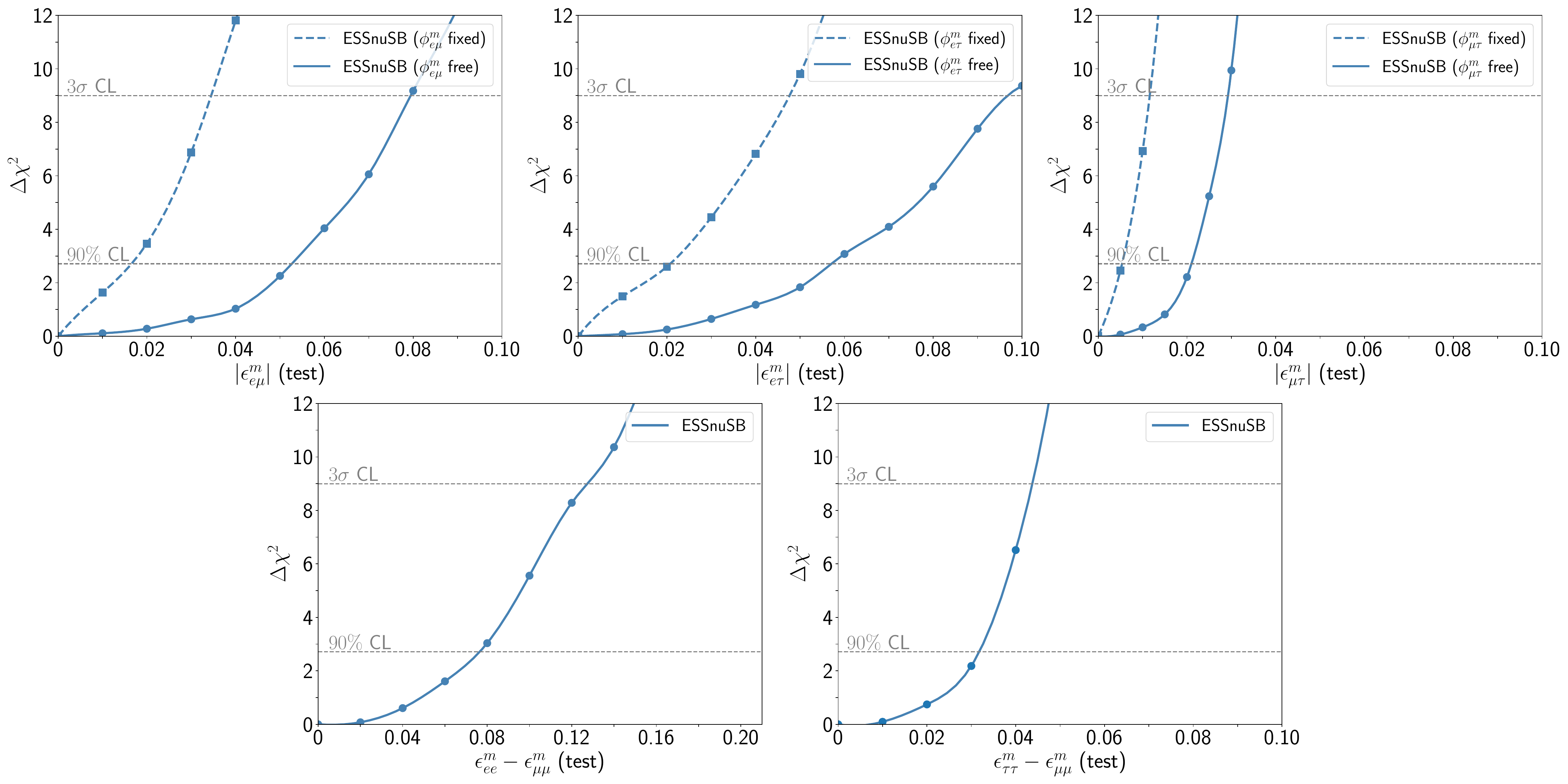}
    \caption{Expected sensitivities to $|\epsilon_{e\mu}^m|, |\epsilon_{e\tau}^m|, |\epsilon_{\mu\tau}^m|, \epsilon_{ee}^m - \epsilon_{\mu\mu}^m$ and $\epsilon_{\tau\tau}^m - \epsilon_{\mu\mu}^m$ for observing atmospheric neutrino oscillations at the ESSnuSB far detector. The simulated data points are fitted with cubic splines. The true neutrino mass ordering is assumed to be normal ordering, while the true values of the neutrino oscillation parameters are assumed to be the same as the global best-fit values~\cite{Esteban:2024eli,NuFIT:6.0}. The dashed lines correspond to the $90\%$ and $3\sigma$~CL limits. Note the chosen value range for $\epsilon_{ee}^m - \epsilon_{\mu\mu}^m$.}
    \label{fig:nsi_bounds}
\end{figure}

In Figure~\ref{fig:nsi_bounds}, individual sensitivities are shown for the matter NSI parameters $\epsilon_{e\mu}^m, \epsilon_{e\tau}^m, \epsilon_{\mu\tau}^m$, and also for $\epsilon_{ee}^m - \epsilon_{\mu\mu}^m$ and $\epsilon_{\tau\tau}^m - \epsilon_{\mu\mu}^m$. The sensitivities are obtained by analyzing the MC event samples for atmospheric neutrinos at ESSnuSB, see Section~\ref{sec:configuration} for further details. In each panel, $\Delta \chi^2 = \chi^2_{\rm NSI} - \chi^2_{\rm SI}$ is presented as a function of the test values for the magnitude of the depicted matter NSI parameter. The true values of the matter NSI parameters are assumed to be zero. Here $\chi^2_{\rm NSI}$ represents the $\chi^2$ value that has been computed with NSI effects, whereas $\chi^2_{\rm SI}$ corresponds to the case where only standard matter effects are applied. The sensitivities are obtained with $5.4~{\rm Mt} \cdot {\rm year}$ exposure. The true neutrino mass ordering is assumed to be normal ordering. The minimization of $\chi_{}^2$ is carried out for both normal ordering and inverted ordering. For the off-diagonal NSI parameters $\epsilon_{e\mu}^m, \epsilon_{e\tau}^m$ and $\epsilon_{\mu\tau}^m$, the phases $\phi_{e\mu}^m, \phi_{e\tau}^m$ and $\phi_{\mu\tau}^m$ are either allowed to vary freely (blue dots) or they are fixed at zero (blue squares) in the minimization of $\chi_{}^2$. The simulated data points are fitted with cubic splines (blue solid and dashed curves). For convenience, the $90\%$ and $3\sigma$~CL limits (1~d.o.f.) are shown by the gray dashed lines. Similar $\Delta \chi^2$ distribution is found for negative $\epsilon_{\tau\tau}^m - \epsilon_{\mu\mu}^m$ values, whereas no significant constraints could be placed on negative values for $\epsilon_{ee}^m - \epsilon_{\mu\mu}^m$.

In the case that no significant deviations from the SM are found, the atmospheric neutrino data collected in the ESSnuSB far detector could be used to set upper bounds on the effective matter NSI parameters. In Figure~\ref{fig:nsi_bounds}, the expected upper bounds can be obtained at the points where the blue solid curves and blue dashed curves intersect with the gray dashed lines. In Table~\ref{tab:nsi_bounds}, the expected upper bounds are listed for ESSnuSB. The upper bounds have been obtained at $90\%$~CL and $3\sigma$~CL given by $\Delta \chi_{}^2 = 2.71$ and $\Delta \chi_{}^2 = 9$, respectively. The numerical values have been obtained by applying linear interpolation. Note that the upper bounds have been obtained for general NSI couplings, including effects from down quarks, up quarks and electrons. The results obtained at $90\%$ CL for free $\phi_{\alpha\beta}^m$ ($\alpha,\beta = e,\mu,\tau$) suggest that ESSnuSB would be able to improve the current bounds on $\epsilon_{e\mu}^m,\epsilon_{e\tau}^m$ and $\epsilon_{\tau\tau}^m - \epsilon_{\mu\mu}^m$ by approximately factors of 3.2, 3.9 and 1.3, respectively. At the same time, ESSnuSB would be able to constrain $\epsilon_{ee}^m - \epsilon_{\mu\mu}^m$ to an accuracy of several percent at $90\%$ CL. This is a noteworthy result, as no experimental bounds have been placed on $\epsilon_{ee}^m - \epsilon_{\mu\mu}^m$ for atmospheric neutrinos. In contrast, the current experimental constraint on $\epsilon_{\mu\tau}^m$ is more stringent than those that could be obtained for ESSnuSB. The comparison between the current constraints and those obtained for ESSnuSB has been done by re-scaling the constraints obtained for ESSnuSB to account NSI couplings for down quarks only.

To be even more precise, the currently leading constraints on the NSI parameters have been set to $|\epsilon_{e\mu}^m| < 0.056$ and $|\epsilon_{e\tau}^m| < 0.074$ ($90\%$~CL) by ORCA6~\cite{KM3NeT:2024pte}, $|\epsilon_{\mu\tau}^m| < 0.0041$ by IceCube~\cite{IceCube:2022ubv}, and $|\epsilon_{\tau\tau}^m - \epsilon_{\mu\mu}^m| < 0.014$ by DeepCore~\cite{IceCubeCollaboration:2021euf}. Note that these upper bounds are reported assuming NSI with down quarks only. The upper bound on $\epsilon_{ee}^m - \epsilon_{\tau\tau}^m$ has not been reported by experimental collaborations, but it has been set to $\epsilon_{ee}^m - \epsilon_{\tau\tau}^m < 0.17$ by the global fit results~\cite{Chatterjee:2024ein}. Therefore, ESSnuSB could potentially improve the upper bounds on $|\epsilon_{e\mu}^m|, |\epsilon_{e\tau}^m|$ and $\epsilon_{\tau\tau}^m - \epsilon_{\mu\mu}^m$. In contrast, the upper bound obtained from IceCube on $|\epsilon_{\mu\tau}^m|$ is found to be more stringent than what could be expected for ESSnuSB.

The complex phases $\phi_{e\mu}^m, \phi_{e\tau}^m$ and $\phi_{\mu\tau}^m$ are observed to have a significant effect on the sensitivities to the matter NSI parameters $\epsilon_{e\mu}^m, \epsilon_{e\tau}^m$ and $\epsilon_{\mu\tau}^m$. When the complex phases $\phi_{e\mu}^m$ and $\phi_{e\tau}^m$ are allowed to vary freely, the expected upper bounds for $|\epsilon_{e\mu}^m|$ and $|\epsilon_{e\tau}^m|$ at $90\%$ CL are relaxed by factors of 3.3 and 2.7, respectively. The corresponding effect of $\phi_{\mu\tau}^m$ for $|\epsilon_{\mu\tau}^m|$ would be a reduction by a factor of 4.0.
\begin{table}[!t]
\begin{center}
\begin{tabular}{|l|C{1.8cm}|C{1.8cm}|C{1.8cm}|C{1.8cm}|C{1.8cm}|}\hline
{\bf Upper bound} & {$\bm{|\epsilon_{e\mu}^m|}$} & {$\bm{|\epsilon_{e\tau}^m|}$} & {\bf $\bm{|\epsilon_{\mu\tau}^m|}$} & {$\bm{\epsilon_{ee}^m - \epsilon_{\mu\mu}^m}$} & {$\bm{\epsilon_{\tau\tau}^m - \epsilon_{\mu\mu}^m}$} \\ \hline
ESSnuSB ($3\sigma$~CL, $
\phi_{\alpha\beta}^m$ free) & $0.079$ & $0.098$ & $0.029$ & $0.13$ & $0.043$ \\ \hline
ESSnuSB ($3\sigma$~CL, $\phi_{\alpha\beta}^m$ fixed) & $0.034$ & $0.047$ & $0.011$ & $0.13$ & $0.043$ \\ \hline
ESSnuSB ($90\%$~CL, $\phi_{\alpha\beta}^m$ free) & $0.053$ & $0.057$ & $0.021$ & $0.075$ & $0.031$ \\ \hline
ESSnuSB ($90\%$~CL, $\phi_{\alpha\beta}^m$ fixed) & $0.016$ & $0.021$ & $0.0053$ & $0.075$ & $0.031$ \\ \hline
\end{tabular}
\end{center}
\caption{\label{tab:nsi_bounds}Bounds on the NSI parameters for observing atmospheric neutrino oscillations at the ESSnuSB far detector. The bounds are obtained by applying the Wilks' theorem, while keeping the complex phase $\phi_{\alpha \beta}^m$ either free or fixed in the minimization ($\alpha,\beta = e, \mu, \tau$). The results have been obtained assuming normal ordering to be the true neutrino mass ordering.}
\end{table}

The results reported in Figure~\ref{fig:nsi_bounds} and Table~\ref{tab:nsi_bounds} are obtained using the standard $\chi^2$ fitting methods as described in Ref.~\cite{ESSnuSB:2024wet}. In the following, we additionally discuss the potential impacts of three different sources of uncertainties that are not considered in the standard analysis. They are (i) the systematic uncertainty related to the $\nu_\mu/\nu_e$ ratio in the atmospheric neutrino fluxes, (ii) the mis-identification between electron-like and muon-like events, and (iii) the atmospheric muon background. Since there are currently no precise estimates for the mis-identification probability or the atmospheric muon background at the ESSnuSB far detectors, the investigation for those two uncertainties focuses on considering the maximum impact that they can have on the atmospheric neutrino analysis.

First, the impact of the $\nu_\mu/\nu_e$ uncertainty is investigated. To do this, we have added a new pull-parameter $\zeta_6$ to parametrize the systematic uncertainty for the $\nu_\mu/\nu_e$ ratio, see {\em e.g.} Ref.~\cite{Gonzalez-Garcia:2004pka} for implementation. We have explicitly checked the effect of this systematic uncertainty by assuming a conservative 5\% uncertainty. The impact of this uncertainty on the results reported in Table~\ref{tab:nsi_bounds} is found to be negligible.

Next, we investigate the impact of flavor mis-identification. To do this, mis-identification probabilities between $0.5\%$ and $2.0\%$ are used to study the effect on the results reported in Table~\ref{tab:nsi_bounds}. For $2.0\%$ mis-identification, we find that the upper bound on $|\epsilon_{e\mu}^m|$ changes from $0.079$ ({\em cf.} Table~\ref{tab:nsi_bounds}) to $0.0826$, which corresponds to a change of 4.0\% that constitutes the maximal impact. In the case where the complex phase is fixed, the change is found to be even smaller. From this, we conclude that the changes in the sensitivities are rather small, and hence, this impact is neglected.

Finally, to discuss the impact of the atmospheric muon background, the contribution of the down-going atmospheric neutrino events in the NSI parameter sensitivities is assessed. This is checked by excluding all down-going events from the numerical analysis, {\em i.e.} all electron-like and muon-like events that satisfy $\cos \theta_z > 0$. Excluding the down-going events, the upper bound on $|\epsilon_{e\mu}^m|$ changes from $0.079$ ({\em cf.} Table~\ref{tab:nsi_bounds}) to $0.0830$, and therefore, the expected upper bound increases by 4.4\%. The events corresponding to $\cos \theta_z \in [0, 0.3]$ are found to yield the most significant contribution among the down-going events. This observation confirms that the sensitivities to NSI parameters arise mainly from up-going events, which indicates that the impact of the atmospheric muon background is of minor importance, and therefore, we have omitted this impact as well.

\section{\label{sec:nsi_effects}Implications on mass ordering and $\theta_{23}$ octant}

In addition to matter NSI parameters, atmospheric neutrino oscillations are sensitive to the neutrino mass ordering and the $\theta_{23}$ octant~\cite{ESSnuSB:2024wet}. However, these sensitivities can potentially be affected by the possibility of having NSI in neutrino propagation. In this section, the impact of the matter NSI parameters $\epsilon_{e\mu}^m, \epsilon_{e\tau}^m, \epsilon_{\mu\tau}^m, \epsilon_{ee}^m - \epsilon_{\mu\mu}^m$ and $\epsilon_{\tau\tau}^m - \epsilon_{\mu\mu}^m$ is examined for the sensitivities to the neutrino mass ordering and the $\theta_{23}^{}$ octant for atmospheric neutrinos at the ESSnuSB far detector.

\begin{figure}[!t]
    \centering
    \includegraphics[width=0.95\linewidth]{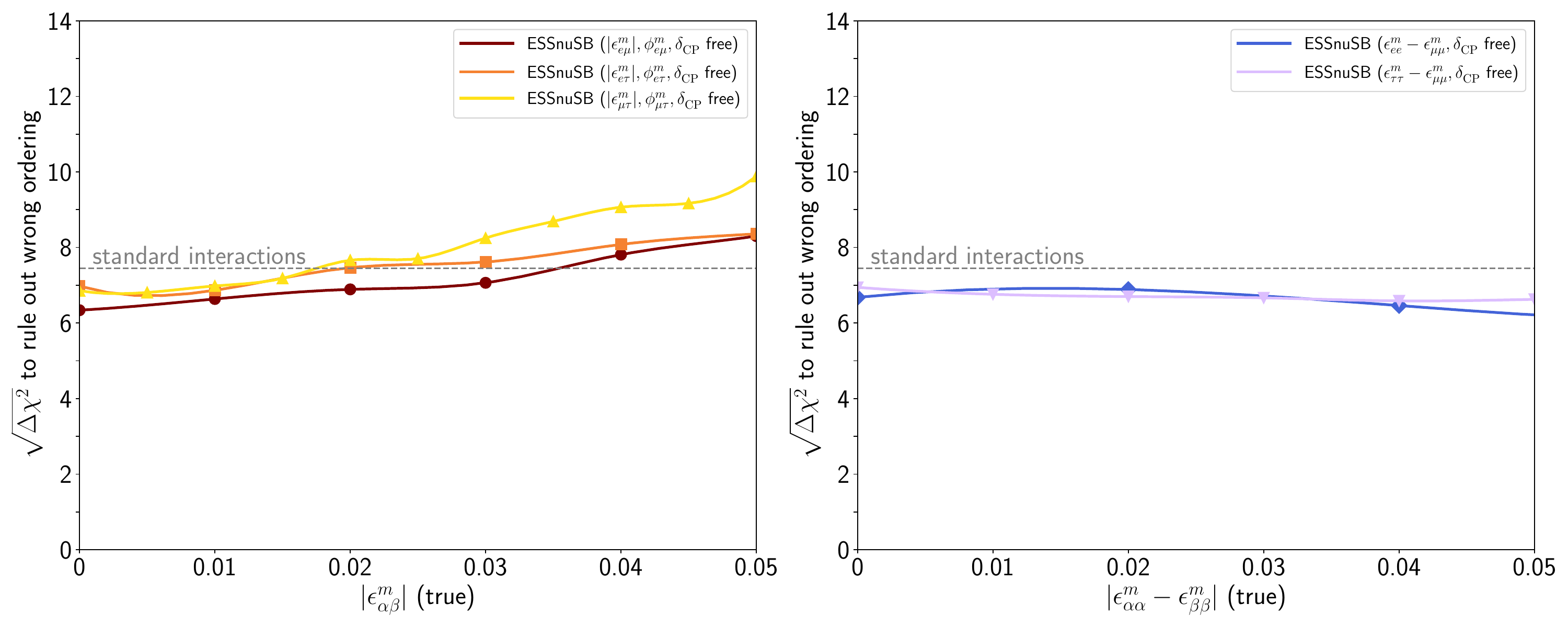}
    \caption{Sensitivity to neutrino mass ordering in presence of matter NSI parameters. For each true value, $\chi^2$ is minimized over the test values of the given matter NSI parameter in addition to the standard parameters. The true neutrino mass ordering is assumed to be normal ordering.}
    \label{fig:mass_ordering}
\end{figure}

In Figure~\ref{fig:mass_ordering}, the sensitivities to neutrino mass ordering in presence of matter NSI are presented for one matter NSI parameter at a time. The $\sqrt{\Delta\chi_{}^2}$ distributions indicate the number of standard deviations at which inverted neutrino mass ordering can be ruled out for each matter NSI parameter. Considered are the matter NSI parameters $\epsilon_{e\mu}^m$ (circles), $\epsilon_{e\tau}^m$ (squares), $\epsilon_{\mu\tau}^m$ (triangles), $\epsilon_{ee}^m - \epsilon_{\mu\mu}^m$ (diamonds) and $\epsilon_{\tau\tau}^m - \epsilon_{\mu\mu}^m$ (inverted triangles). The computed sensitivities are fitted with splines (solid curves). The sensitivity to the neutrino mass ordering is computed as $\sqrt{\Delta \chi^2} = \sqrt{\chi_{\rm IO}^2 - \chi_{\rm NO}^2}$, where $\chi_{\rm IO}^2$ is minimized for the $\Delta m_{31}^2$ values that correspond to inverted ordering, $\Delta m_{31}^2 < 0$, and $\chi_{\rm NO}^2$ is minimized for the values matching normal ordering, $\Delta m_{31}^2 > 0$, respectively. The corresponding sensitivity is also presented without matter NSI parameters (gray dashed line), in which case $\delta_{\rm CP}$ and the relevant matter NSI parameters are fixed at zero. We have furthermore assumed the true values of $\phi_{e\mu}^m, \phi_{e\tau}^m$ and $\phi_{\mu\tau}^m$ to be zero. In the left panel, the sensitivities are shown as functions of the true value of the magnitude $|\epsilon_{\alpha \beta}^m|$, where $\alpha, \beta = e, \mu, \tau$. For those sensitivities, the $\chi^2$ function is minimized for the standard neutrino oscillation parameters $\theta_{23}^{}, \Delta m_{31}^2$ and $ \delta_{CP}^{}$, and also for the non-standard parameters $|\epsilon_{\alpha\beta}^m|$ and $\phi_{\alpha\beta}^m$. The remaining matter NSI parameters are fixed at zero. In the right panel, the sensitivities to the neutrino mass ordering are shown for the diagonal parameters $|\epsilon_{\alpha\alpha}^m - \epsilon_{\beta\beta}^m|$. For those matter NSI parameters, the minimization of $\chi_{}^2$ includes $\epsilon_{\alpha\alpha}^m - \epsilon_{\beta\beta}^m$ in addition to the standard neutrino oscillation parameters. One can expect analogous results for the case where the true neutrino mass ordering is assumed to be inverted ordering.

When the matter NSI effects are taken into account, the sensitivity to the mass ordering is expected to decrease as a result of the increased number of free parameters. This can be observed in Figure~\ref{fig:mass_ordering}, where the sensitivity to exclude inverted ordering is reduced from the standard interactions case for the true values $|\epsilon_{\alpha \beta}^m| = 0$ ($|\epsilon_{\alpha \alpha}^m - \epsilon_{\beta \beta}^m| = 0$) for the off-diagonal (diagonal) matter NSI parameters. For $|\epsilon_{e\mu}^m|,|\epsilon_{e\tau}^m|$ and $|\epsilon_{\mu\tau}^m|$, the sensitivity generally increases as the true value increases. In contrast, the sensitivities to rule out the wrong mass ordering for $|\epsilon_{ee}^m - \epsilon_{\mu\mu}^m|$ or $|\epsilon_{\tau\tau}^m - \epsilon_{\mu\mu}^m|$ are mostly independent of their corresponding true values. Figure~\ref{fig:mass_ordering} shows that taking the magnitude and phase of $\epsilon_{e\mu}^m$ as free parameters results in approximately $0.6\sigma-1.3\sigma$~CL decrease in the mass ordering sensitivity. The ESSnuSB setup can be expected to exclude the wrong neutrino mass ordering by about $6.1\sigma$~CL or better. Note that the matter NSI parameters $\epsilon_{ee}^m -\epsilon_{\mu\mu}^m$ and $\epsilon_{\tau\tau}^m -\epsilon_{\mu\mu}^m$ can have negative values.

\begin{figure}[!t]
    \centering
    \includegraphics[width=0.95\linewidth]{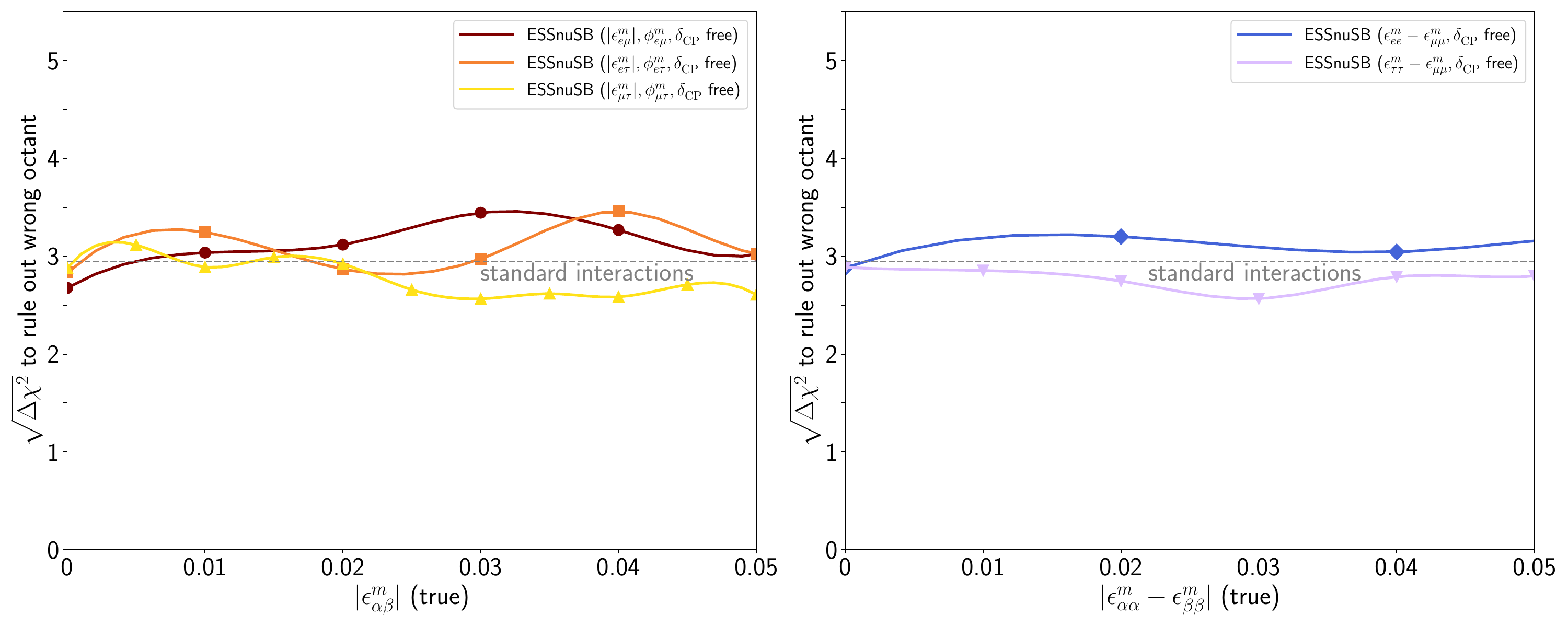}
    \caption{Effect of matter NSI parameters $|\epsilon_{e\mu}^m|, |\epsilon_{e\tau}^m|, \epsilon_{\mu\tau}^m|, |\epsilon_{ee}^m - \epsilon_{\mu\mu}^m|$ and $|\epsilon_{\tau\tau}^m - \epsilon_{\mu\mu}^m|$ on the sensitivity to the octant of $\theta_{23}^{}$. For each true value, $\chi^2$ is minimized over the test values of the given matter NSI parameter in addition to the standard parameters. The true neutrino mass ordering is assumed to be normal ordering.}
    \label{fig:th23_octant}
\end{figure}

The effect of matter NSI parameters can similarly be investigated for the determination of the $\theta_{23}^{}$ octant. In Figure~\ref{fig:th23_octant}, the sensitivity to the octant of $\theta_{23}^{}$ is shown as a function of the magnitudes of the effective matter NSI parameters. As before, the effect is studied for only one matter NSI parameter at a time. The true neutrino mass ordering is assumed to be normal ordering. The true values of $\phi_{e\mu}^m, \phi_{e\tau}^m$ and $\phi_{\mu\tau}^m$ are again assumed to be zero. The sensitivity to the octant of $\theta_{23}^{}$ is obtained by minimizing the $\chi_{}^2$ function for $\sqrt{\Delta\chi_{}^2} = \sqrt{\chi_{\rm HO}^2 - \chi_{\rm LO}^2}$, where HO and LO denote the high octant and the low octant, respectively. Here $\chi_{\rm HO}^2$ is obtained by minimizing $\chi_{}^2$ for the high octant values, {\em i.e.}, $\sin_{}^2\theta_{23}^{} > 0.5$, while $\chi_{\rm LO}^2$ is obtained by minimizing $\chi_{}^2$ for the low octant values, in which case $\sin_{}^2\theta_{23}^{} < 0.5$. The left panel shows the sensitivities to $\theta_{23}^{}$ octant for the off-diagonal matter NSI parameters $\epsilon_{\alpha\beta}^m$, whereas the effects of the diagonal matter NSI parameters $\epsilon_{\alpha\alpha}^m - \epsilon_{\beta\beta}^m$ are illustrated in the right panel. The sensitivities that are obtained for standard interactions without matter NSI effects are depicted with the gray dashed lines. 

Figure~\ref{fig:th23_octant} reveals that matter NSI could have a noticeable effect on the sensitivity to the octant of $\theta_{23}^{}$. For the true values $|\epsilon_{\alpha\beta}^m| = 0$ and $|\epsilon_{\alpha\alpha}^m - \epsilon_{\beta\beta}^m| = 0$, the sensitivities to exclude the wrong octant would be lowered by $0.1\sigma-0.4\sigma$~CL, while the minimum sensitivity would be $2.6\sigma$~CL. Furthermore, it is shown in Figure~\ref{fig:th23_octant} that for certain values of $|\epsilon_{\alpha\beta}^m|$ and $|\epsilon_{\alpha\alpha}^m - \epsilon_{\beta\beta}^m|$ the sensitivity to the octant of $\theta_{23}^{}$ could be higher than what would be obtained for the case where all matter NSI parameters as well as $\delta_{\rm CP}^{}$ are fixed at zero. Only $\epsilon_{\tau\tau}^m - \epsilon_{\mu\mu}^m$ is an exception, as the sensitivity to the octant of $\theta_{23}^{}$ would remain consistently below the dashed gray line regardless of the true value of $|\epsilon_{\tau\tau}^m - \epsilon_{\mu\mu}^m|$. ESSnuSB would therefore retain a noteworthy sensitivity for the octant of $\theta_{23}^{}$ despite the increased number of free parameters.

In conclusion, the atmospheric neutrino data that could be collected in the ESSnuSB far detector would be sufficient to determine the neutrino mass ordering and provide sensitivity to the octant of $\theta_{23}$. Especially, the sensitivity to the mass ordering would be well above the $5\sigma$~CL discovery limit even in the case that one of the matter NSI parameters is allowed to run free in the statistical analysis.

\section{\label{sec:concl}Summary and conclusions}

In this work, the phenomenological implications of matter NSI parameters were considered for atmospheric neutrinos at the ESSnuSB far detector. The sensitivities were derived for the effective matter NSI parameters $\epsilon_{e\mu}^m, \epsilon_{e\tau}^m, \epsilon_{\mu\tau}^m, \epsilon_{ee}^m - \epsilon_{\mu\mu}^m$ and $\epsilon_{\tau\tau}^m - \epsilon_{\mu\mu}^m$ by analyzing Monte Carlo events that were generated for the ESSnuSB far detector assuming $5.4~{\rm Mt}\cdot{\rm year}$ exposure. In the case where no signatures of new physics were observed, the atmospheric neutrino data expected for ESSnuSB would be able to place the upper bounds $|\epsilon_{e\mu}^m| < 0.053, |\epsilon_{e\tau}^m| < 0.057$ and $|\epsilon_{\mu\tau}^m| < 0.021$ on the complex matter NSI parameters at $90\%$~CL, assuming the true neutrino mass ordering to be normal ordering. The effect of the complex phases $\phi_{e\mu}^m, \phi_{e\tau}^m$ and $\phi_{\mu\tau}^m$ was taken into account in the determination of the sensitivities. Similarly, the diagonal matter NSI parameters $\epsilon_{ee}^m - \epsilon_{\mu\mu}^m$ and $\epsilon_{\tau\tau}^m - \epsilon_{\mu\mu}^m$ could be constrained to $\epsilon_{ee}^m - \epsilon_{\mu\mu}^m < 0.075$ and $\epsilon_{\tau\tau}^m - \epsilon_{\mu\mu}^m < 0.031$ for positive values. A similar constraint could be obtained for negative values of $\epsilon_{\tau\tau}^m - \epsilon_{\mu\mu}^m$, whereas no significant bound could be placed on $\epsilon_{ee}^m - \epsilon_{\mu\mu}^m$ for its negative values. In the case that the complex phases $\phi_{e\mu}^m, \phi_{e\tau}^m$ and $\phi_{\mu\tau}^m$ are fixed at zero, the expected upper bounds for $\epsilon_{e\mu}^m, \epsilon_{e\tau}^m$ and $\epsilon_{\mu\tau}^m$ would be $|\epsilon_{e\mu}^m| < 0.016, |\epsilon_{e\tau}^m| < 0.021$ and $|\epsilon_{\mu\tau}^m| < 0.0053$.

The effects of the matter NSI parameters were additionally investigated for determining the neutrino mass ordering and the octant of $\theta_{23}^{}$. In the case that one of the matter NSI parameters were treated as a free parameter, the sensitivity to rule out the wrong mass ordering would remain approximately at $6.1\sigma$~CL or higher. Similarly, letting the matter NSI parameters to run free in the determination of the octant of $\theta_{23}^{}$ would keep the sensitivity at $2.6\sigma$~CL or higher. Therefore, the atmospheric neutrino data expected for ESSnuSB would be sufficient to resolve the neutrino mass ordering and provide sensitivities to both the matter NSI parameters and the octant of $\theta_{23}^{}$. Such results would be complementary to the main experimental program of the ESSnuSB project.

\section*{Acknowledgments}

Funded by the European Union, Project 101094628. Views and opinions expressed are however those of the author(s) only and do not necessarily reflect those of the European Union. Neither the European Union nor the granting authority can be held responsible for them. We acknowledge further support provided by the following research funding agencies: Centre National de la Recherche Scientifique, France; Deutsche Forschungsgemeinschaft Projektnummer 423761110, and under the Excellence Strategy of the Federal Government and the L{\"a}nder, Germany; Ministry of Science and Education of Republic of Croatia grant No. KK.01.1.1.01.0001; the European Union’s Horizon 2020 research and innovation programme under the Marie Sk{\l}odowska-Curie grant agreement No 860881-HIDDeN; the European Union NextGenerationEU, through the National Recovery and Resilience Plan of the Republic of Bulgaria, project No. BG-RRP-2.004-0008-C01; Roland Gustafssons Stiftelse f\"or teoretisk fysik, Sweden; as well as support provided by the universities and laboratories to which the authors of this report are affiliated, see the author list on the first page. The authors are grateful for the computing resources that were provided by the Division of Condensed Matter Theory at KTH Royal Institute of Technology.

\appendix

\section{\label{Appendix:Events}Additional investigations with atmospheric neutrino events}

In this appendix, the effect in the expected atmospheric neutrino events is investigated for the matter NSI parameters $\epsilon_{e\tau}^m, \epsilon_{\mu\tau}^m, \epsilon_{ee}^m - \epsilon_{\mu\mu}^m$ and $\epsilon_{\tau\tau}^m - \epsilon_{\mu\mu}^m$. For the off-diagonal parameter $\epsilon_{e\mu}^m$, we refer the reader to Figure~\ref{fig:nsi_events_emu} in Section~\ref{sec:nsi_constraints}. It is also shown in this appendix how a single $1\sigma$ pull for the zenith angle dependence error affects the electron-like and muon-like events in the ESSnuSB far detectors. In each case, atmospheric neutrino events are compared for 10~years of data taking.

\begin{figure}[!t]
    \centering
    \includegraphics[width=1.0\linewidth]{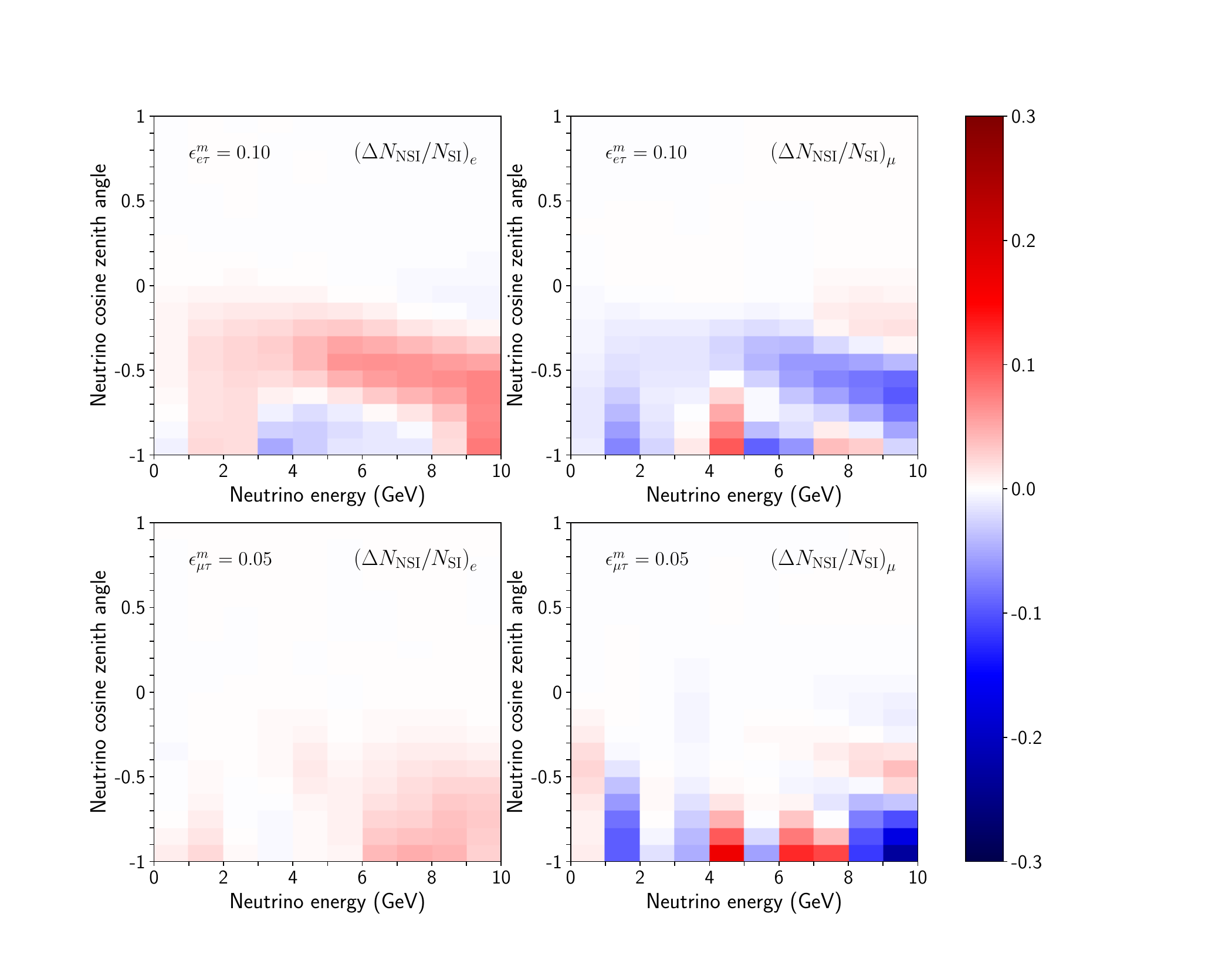}
    \caption{Same as in Figure~\ref{fig:nsi_events_emu}, but for $\epsilon_{e\tau}^m = 0.10$ and $\epsilon_{\mu\tau}^m = 0.05$.}
    \label{fig:nsi_events_off-diagonal}
\end{figure}

In Figure~\ref{fig:nsi_events_off-diagonal}, the relative differences in the atmospheric neutrino events are shown for the off-diagonal parameters $\epsilon_{e\tau}^m$ and $\epsilon_{\mu\tau}^m$. In the top-left and top-right panels, the relative differences $\Delta N_{\rm NSI} / N_{\rm SI} = (N_{\rm NSI} - N_{\rm SI}) / N_{\rm SI}$ are shown for the electron-like and muon-like samples, respectively, when $N_{\rm NSI}$ is computed for $\epsilon_{e\tau}^m = 0.10$. In a similar manner, in the bottom-left and bottom-right panels, the relative differences are shown for the electron-like and muon-like events in the case where $\epsilon_{\mu\tau}^m = 0.05$. Note that the relative differences are shown up to $10$~GeV neutrino energy. 

\begin{figure}[!t]
    \centering
    \includegraphics[width=1.0\linewidth]{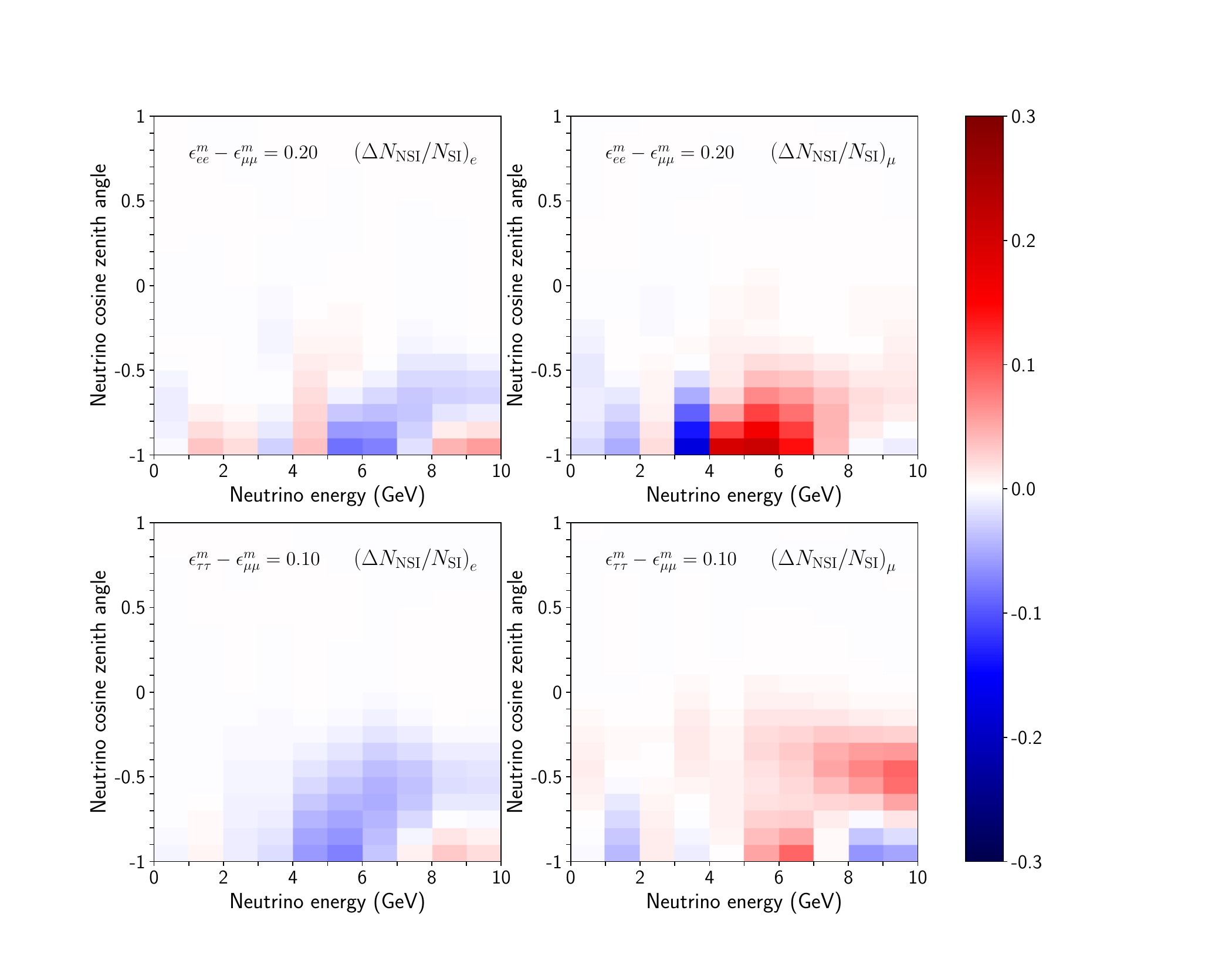}
    \caption{Same as in Figure~\ref{fig:nsi_events_emu}, but for $\epsilon_{ee}^m - \epsilon_{\mu\mu}^m = 0.20$ and $\epsilon_{\tau\tau}^m - \epsilon_{\mu\mu}^m = 0.10$.}
    \label{fig:nsi_events_on_diagonal}
\end{figure}

In Figure~\ref{fig:nsi_events_on_diagonal}, the relative differences in the expected atmospheric neutrino events are shown for the diagonal parameters $\epsilon_{ee}^m - \epsilon_{\mu\mu}^m$ and $\epsilon_{\tau\tau}^m - \epsilon_{\mu\mu}^m$. In this case, the relative differences in electron-like and muon-like events are presented for $\epsilon_{ee}^m - \epsilon_{\mu\mu}^m = 0.20$ in top-left and top-right panels, and for $\epsilon_{\tau\tau}^m - \epsilon_{\mu\mu}^m = 0.10$ in bottom-left and bottom-right panels, respectively. For consistency, the relative differences are shown up to $10$~GeV neutrino energy.

\begin{figure}[!h]
    \centering
    \includegraphics[width=0.9\linewidth]{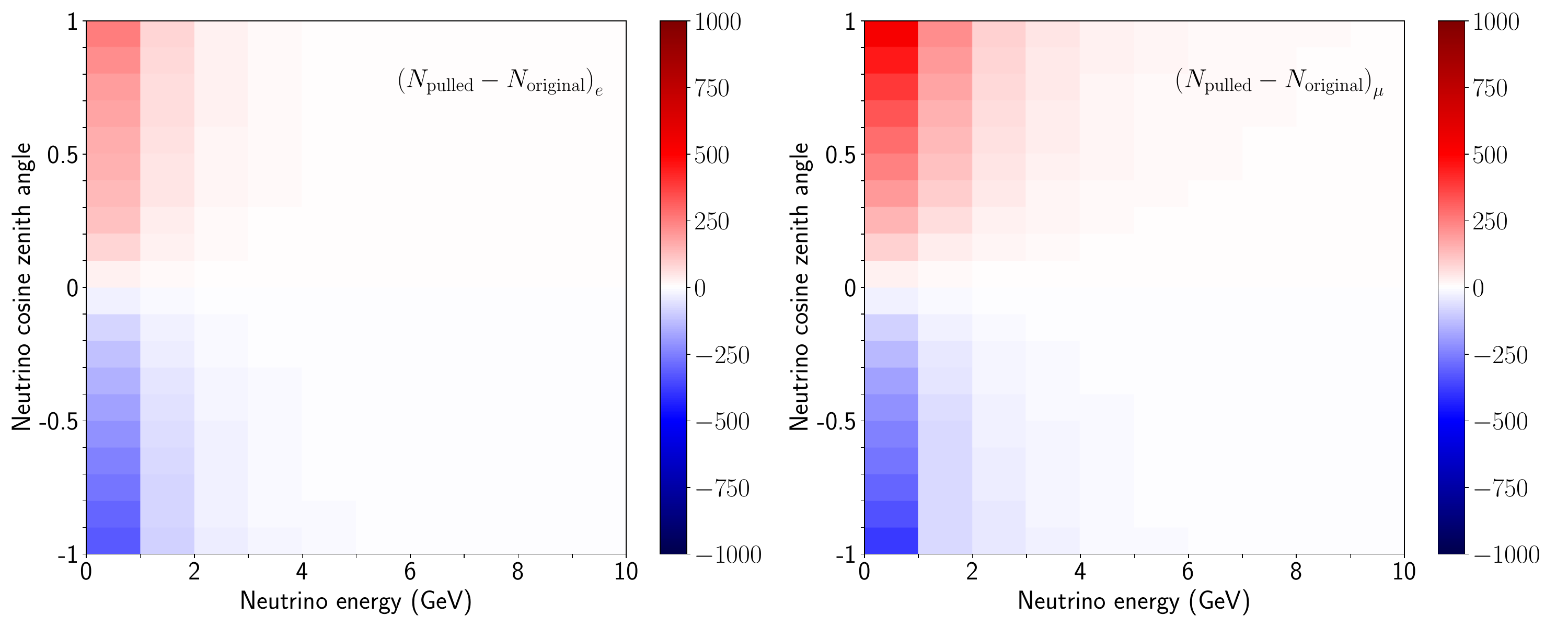}
    \caption{Expected differences in atmospheric neutrino events after performing systematic pulls on $e$-like events (left panel) and $\mu$-like events (right panel) for the zenith angle dependence error. The corresponding pull-parameters are varied within their $1\sigma$ limits. Here $N_{\rm pulled}$ and $N_{\rm original}$ refer to the pulled and original events, respectively. All NSI parameters are set to zero.}
    \label{fig:pulled_vs_original}
\end{figure}

In Figure~\ref{fig:pulled_vs_original}, the effect of the zenith angle dependence systematic uncertainty is shown for the electron-like and muon-like events. The pulled events $N_{\rm pulled}$ correspond to the atmospheric neutrino event spectrum in the case where a systematic pull has been performed on the original atmospheric neutrino events $N_{\rm original}$ at $1\sigma$~CL for the cosine zenith angle dependence error. This is achieved by setting the relevant pull-parameter to $\zeta_3 = 1$. Other pull-parameters are set to zero. The figure shows that the effect on the expected atmospheric neutrino events is the most significant near $\cos \theta_z = \pm 1$. Correspondingly, the effect of the zenith angle dependence error is the least significant near $\cos \theta_z = 0$.

\bibliographystyle{apsrev4-2}
\bibliography{references}

\begin{thebibliography}{56}%
\makeatletter
\providecommand \@ifxundefined [1]{%
 \@ifx{#1\undefined}
}%
\providecommand \@ifnum [1]{%
 \ifnum #1\expandafter \@firstoftwo
 \else \expandafter \@secondoftwo
 \fi
}%
\providecommand \@ifx [1]{%
 \ifx #1\expandafter \@firstoftwo
 \else \expandafter \@secondoftwo
 \fi
}%
\providecommand \natexlab [1]{#1}%
\providecommand \enquote  [1]{``#1''}%
\providecommand \bibnamefont  [1]{#1}%
\providecommand \bibfnamefont [1]{#1}%
\providecommand \citenamefont [1]{#1}%
\providecommand \href@noop [0]{\@secondoftwo}%
\providecommand \href [0]{\begingroup \@sanitize@url \@href}%
\providecommand \@href[1]{\@@startlink{#1}\@@href}%
\providecommand \@@href[1]{\endgroup#1\@@endlink}%
\providecommand \@sanitize@url [0]{\catcode `\\12\catcode `\$12\catcode
  `\&12\catcode `\#12\catcode `\^12\catcode `\_12\catcode `\%12\relax}%
\providecommand \@@startlink[1]{}%
\providecommand \@@endlink[0]{}%
\providecommand \url  [0]{\begingroup\@sanitize@url \@url }%
\providecommand \@url [1]{\endgroup\@href {#1}{\urlprefix }}%
\providecommand \urlprefix  [0]{URL }%
\providecommand \Eprint [0]{\href }%
\providecommand \doibase [0]{https://doi.org/}%
\providecommand \selectlanguage [0]{\@gobble}%
\providecommand \bibinfo  [0]{\@secondoftwo}%
\providecommand \bibfield  [0]{\@secondoftwo}%
\providecommand \translation [1]{[#1]}%
\providecommand \BibitemOpen [0]{}%
\providecommand \bibitemStop [0]{}%
\providecommand \bibitemNoStop [0]{.\EOS\space}%
\providecommand \EOS [0]{\spacefactor3000\relax}%
\providecommand \BibitemShut  [1]{\csname bibitem#1\endcsname}%
\let\auto@bib@innerbib\@empty
\bibitem [{\citenamefont {Esteban}\ \emph
  {et~al.}(2024{\natexlab{a}})\citenamefont {Esteban}, \citenamefont
  {Gonzalez-Garcia}, \citenamefont {Maltoni}, \citenamefont {Martinez-Soler},
  \citenamefont {Pinheiro},\ and\ \citenamefont {Schwetz}}]{Esteban:2024eli}%
  \BibitemOpen
  \bibfield  {author} {\bibinfo {author} {\bibfnamefont {I.}~\bibnamefont
  {Esteban}}, \bibinfo {author} {\bibfnamefont {M.~C.}\ \bibnamefont
  {Gonzalez-Garcia}}, \bibinfo {author} {\bibfnamefont {M.}~\bibnamefont
  {Maltoni}}, \bibinfo {author} {\bibfnamefont {I.}~\bibnamefont
  {Martinez-Soler}}, \bibinfo {author} {\bibfnamefont {J.~a.~P.}\ \bibnamefont
  {Pinheiro}},\ and\ \bibinfo {author} {\bibfnamefont {T.}~\bibnamefont
  {Schwetz}},\ }\href {https://doi.org/10.1007/JHEP12(2024)216} {\bibfield
  {journal} {\bibinfo  {journal} {JHEP}\ }\textbf {\bibinfo {volume} {12}},\
  \bibinfo {pages} {216}},\ \Eprint {https://arxiv.org/abs/2410.05380}
  {arXiv:2410.05380 [hep-ph]} \BibitemShut {NoStop}%
\bibitem [{\citenamefont {Ohlsson}(2013)}]{Ohlsson:2012kf}%
  \BibitemOpen
  \bibfield  {author} {\bibinfo {author} {\bibfnamefont {T.}~\bibnamefont
  {Ohlsson}},\ }\href {https://doi.org/10.1088/0034-4885/76/4/044201}
  {\bibfield  {journal} {\bibinfo  {journal} {Rept. Prog. Phys.}\ }\textbf
  {\bibinfo {volume} {76}},\ \bibinfo {pages} {044201} (\bibinfo {year}
  {2013})},\ \Eprint {https://arxiv.org/abs/1209.2710} {arXiv:1209.2710
  [hep-ph]} \BibitemShut {NoStop}%
\bibitem [{\citenamefont {Farzan}\ and\ \citenamefont
  {T\'ortola}(2018)}]{Farzan:2017xzy}%
  \BibitemOpen
  \bibfield  {author} {\bibinfo {author} {\bibfnamefont {Y.}~\bibnamefont
  {Farzan}}\ and\ \bibinfo {author} {\bibfnamefont {M.}~\bibnamefont
  {T\'ortola}},\ }\href {https://doi.org/10.3389/fphy.2018.00010} {\bibfield
  {journal} {\bibinfo  {journal} {Front. in Phys.}\ }\textbf {\bibinfo {volume}
  {6}},\ \bibinfo {pages} {10} (\bibinfo {year} {2018})},\ \Eprint
  {https://arxiv.org/abs/1710.09360} {arXiv:1710.09360 [hep-ph]} \BibitemShut
  {NoStop}%
\bibitem [{\citenamefont {Ge}\ and\ \citenamefont {Parke}(2019)}]{Ge:2018uhz}%
  \BibitemOpen
  \bibfield  {author} {\bibinfo {author} {\bibfnamefont {S.-F.}\ \bibnamefont
  {Ge}}\ and\ \bibinfo {author} {\bibfnamefont {S.~J.}\ \bibnamefont {Parke}},\
  }\href {https://doi.org/10.1103/PhysRevLett.122.211801} {\bibfield  {journal}
  {\bibinfo  {journal} {Phys. Rev. Lett.}\ }\textbf {\bibinfo {volume} {122}},\
  \bibinfo {pages} {211801} (\bibinfo {year} {2019})},\ \Eprint
  {https://arxiv.org/abs/1812.08376} {arXiv:1812.08376 [hep-ph]} \BibitemShut
  {NoStop}%
\bibitem [{\citenamefont {Alekou}\ \emph {et~al.}(2022)\citenamefont {Alekou}
  \emph {et~al.}}]{Alekou:2022emd}%
  \BibitemOpen
  \bibfield  {author} {\bibinfo {author} {\bibfnamefont {A.}~\bibnamefont
  {Alekou}} \emph {et~al.} (\bibinfo {collaboration} {ESSnuSB}),\ }\href
  {https://doi.org/10.1140/epjs/s11734-022-00664-w} {\bibfield  {journal}
  {\bibinfo  {journal} {Eur. Phys. J. ST}\ }\textbf {\bibinfo {volume} {231}},\
  \bibinfo {pages} {3779} (\bibinfo {year} {2022})},\ \bibinfo {note}
  {[Erratum: Eur.Phys.J.ST 232, 15--16 (2023)]},\ \Eprint
  {https://arxiv.org/abs/2206.01208} {arXiv:2206.01208 [hep-ex]} \BibitemShut
  {NoStop}%
\bibitem [{\citenamefont {Abele}\ \emph {et~al.}(2023)\citenamefont {Abele}
  \emph {et~al.}}]{Abele:2022iml}%
  \BibitemOpen
  \bibfield  {author} {\bibinfo {author} {\bibfnamefont {H.}~\bibnamefont
  {Abele}} \emph {et~al.},\ }\href
  {https://doi.org/10.1016/j.physrep.2023.06.001} {\bibfield  {journal}
  {\bibinfo  {journal} {Phys. Rept.}\ }\textbf {\bibinfo {volume} {1023}},\
  \bibinfo {pages} {1} (\bibinfo {year} {2023})},\ \Eprint
  {https://arxiv.org/abs/2211.10396} {arXiv:2211.10396 [physics.ins-det]}
  \BibitemShut {NoStop}%
\bibitem [{\citenamefont {Alekou}\ \emph {et~al.}(2023)\citenamefont {Alekou}
  \emph {et~al.}}]{ESSnuSB:2023ogw}%
  \BibitemOpen
  \bibfield  {author} {\bibinfo {author} {\bibfnamefont {A.}~\bibnamefont
  {Alekou}} \emph {et~al.} (\bibinfo {collaboration} {ESSnuSB}),\ }\href
  {https://doi.org/10.3390/universe9080347} {\bibfield  {journal} {\bibinfo
  {journal} {Universe}\ }\textbf {\bibinfo {volume} {9}},\ \bibinfo {pages}
  {347} (\bibinfo {year} {2023})},\ \Eprint {https://arxiv.org/abs/2303.17356}
  {arXiv:2303.17356 [hep-ex]} \BibitemShut {NoStop}%
\bibitem [{\citenamefont {Baussan}\ \emph {et~al.}(2014)\citenamefont {Baussan}
  \emph {et~al.}}]{ESSnuSB:2013dql}%
  \BibitemOpen
  \bibfield  {author} {\bibinfo {author} {\bibfnamefont {E.}~\bibnamefont
  {Baussan}} \emph {et~al.} (\bibinfo {collaboration} {ESSnuSB}),\ }\href
  {https://doi.org/10.1016/j.nuclphysb.2014.05.016} {\bibfield  {journal}
  {\bibinfo  {journal} {Nucl. Phys. B}\ }\textbf {\bibinfo {volume} {885}},\
  \bibinfo {pages} {127} (\bibinfo {year} {2014})},\ \Eprint
  {https://arxiv.org/abs/1309.7022} {arXiv:1309.7022 [hep-ex]} \BibitemShut
  {NoStop}%
\bibitem [{\citenamefont {Blennow}\ \emph {et~al.}(2015)\citenamefont
  {Blennow}, \citenamefont {Choubey}, \citenamefont {Ohlsson},\ and\
  \citenamefont {Raut}}]{Blennow:2015nxa}%
  \BibitemOpen
  \bibfield  {author} {\bibinfo {author} {\bibfnamefont {M.}~\bibnamefont
  {Blennow}}, \bibinfo {author} {\bibfnamefont {S.}~\bibnamefont {Choubey}},
  \bibinfo {author} {\bibfnamefont {T.}~\bibnamefont {Ohlsson}},\ and\ \bibinfo
  {author} {\bibfnamefont {S.~K.}\ \bibnamefont {Raut}},\ }\href
  {https://doi.org/10.1007/JHEP09(2015)096} {\bibfield  {journal} {\bibinfo
  {journal} {JHEP}\ }\textbf {\bibinfo {volume} {09}},\ \bibinfo {pages}
  {096}},\ \Eprint {https://arxiv.org/abs/1507.02868} {arXiv:1507.02868
  [hep-ph]} \BibitemShut {NoStop}%
\bibitem [{\citenamefont {Delgadillo}\ and\ \citenamefont
  {Miranda}(2023)}]{Delgadillo:2023lyp}%
  \BibitemOpen
  \bibfield  {author} {\bibinfo {author} {\bibfnamefont {L.~A.}\ \bibnamefont
  {Delgadillo}}\ and\ \bibinfo {author} {\bibfnamefont {O.~G.}\ \bibnamefont
  {Miranda}},\ }\href {https://doi.org/10.1103/PhysRevD.108.095024} {\bibfield
  {journal} {\bibinfo  {journal} {Phys. Rev. D}\ }\textbf {\bibinfo {volume}
  {108}},\ \bibinfo {pages} {095024} (\bibinfo {year} {2023})},\ \Eprint
  {https://arxiv.org/abs/2304.05545} {arXiv:2304.05545 [hep-ph]} \BibitemShut
  {NoStop}%
\bibitem [{\citenamefont {Aguilar}\ \emph
  {et~al.}(2024{\natexlab{a}})\citenamefont {Aguilar} \emph
  {et~al.}}]{ESSnuSB:2023lbg}%
  \BibitemOpen
  \bibfield  {author} {\bibinfo {author} {\bibfnamefont {J.}~\bibnamefont
  {Aguilar}} \emph {et~al.} (\bibinfo {collaboration} {ESSnuSB}),\ }\href
  {https://doi.org/10.1103/PhysRevD.109.115010} {\bibfield  {journal} {\bibinfo
   {journal} {Phys. Rev. D}\ }\textbf {\bibinfo {volume} {109}},\ \bibinfo
  {pages} {115010} (\bibinfo {year} {2024}{\natexlab{a}})},\ \Eprint
  {https://arxiv.org/abs/2310.10749} {arXiv:2310.10749 [hep-ex]} \BibitemShut
  {NoStop}%
\bibitem [{\citenamefont {Baxter}\ \emph {et~al.}(2020)\citenamefont {Baxter}
  \emph {et~al.}}]{Baxter:2019mcx}%
  \BibitemOpen
  \bibfield  {author} {\bibinfo {author} {\bibfnamefont {D.}~\bibnamefont
  {Baxter}} \emph {et~al.},\ }\href {https://doi.org/10.1007/JHEP02(2020)123}
  {\bibfield  {journal} {\bibinfo  {journal} {JHEP}\ }\textbf {\bibinfo
  {volume} {02}},\ \bibinfo {pages} {123}},\ \Eprint
  {https://arxiv.org/abs/1911.00762} {arXiv:1911.00762 [physics.ins-det]}
  \BibitemShut {NoStop}%
\bibitem [{\citenamefont {Chatterjee}\ \emph {et~al.}(2023)\citenamefont
  {Chatterjee}, \citenamefont {Lavignac}, \citenamefont {Miranda},\ and\
  \citenamefont {Sanchez~Garcia}}]{Chatterjee:2022mmu}%
  \BibitemOpen
  \bibfield  {author} {\bibinfo {author} {\bibfnamefont {S.~S.}\ \bibnamefont
  {Chatterjee}}, \bibinfo {author} {\bibfnamefont {S.}~\bibnamefont
  {Lavignac}}, \bibinfo {author} {\bibfnamefont {O.~G.}\ \bibnamefont
  {Miranda}},\ and\ \bibinfo {author} {\bibfnamefont {G.}~\bibnamefont
  {Sanchez~Garcia}},\ }\href {https://doi.org/10.1103/PhysRevD.107.055019}
  {\bibfield  {journal} {\bibinfo  {journal} {Phys. Rev. D}\ }\textbf {\bibinfo
  {volume} {107}},\ \bibinfo {pages} {055019} (\bibinfo {year} {2023})},\
  \Eprint {https://arxiv.org/abs/2208.11771} {arXiv:2208.11771 [hep-ph]}
  \BibitemShut {NoStop}%
\bibitem [{\citenamefont {Sim\'on}(2024)}]{Simon:2024xwb}%
  \BibitemOpen
  \bibfield  {author} {\bibinfo {author} {\bibfnamefont {A.}~\bibnamefont
  {Sim\'on}} (\bibinfo {collaboration} {$\nu$ESS}),\ }\href
  {https://doi.org/10.22323/1.441.0171} {\bibfield  {journal} {\bibinfo
  {journal} {PoS}\ }\textbf {\bibinfo {volume} {TAUP2023}},\ \bibinfo {pages}
  {171} (\bibinfo {year} {2024})},\ \Eprint {https://arxiv.org/abs/2401.04074}
  {arXiv:2401.04074 [hep-ex]} \BibitemShut {NoStop}%
\bibitem [{\citenamefont {Ghosh}\ \emph {et~al.}(2020)\citenamefont {Ghosh},
  \citenamefont {Ohlsson},\ and\ \citenamefont
  {Rosauro-Alcaraz}}]{Ghosh:2019zvl}%
  \BibitemOpen
  \bibfield  {author} {\bibinfo {author} {\bibfnamefont {M.}~\bibnamefont
  {Ghosh}}, \bibinfo {author} {\bibfnamefont {T.}~\bibnamefont {Ohlsson}},\
  and\ \bibinfo {author} {\bibfnamefont {S.}~\bibnamefont {Rosauro-Alcaraz}},\
  }\href {https://doi.org/10.1007/JHEP03(2020)026} {\bibfield  {journal}
  {\bibinfo  {journal} {JHEP}\ }\textbf {\bibinfo {volume} {03}},\ \bibinfo
  {pages} {026}},\ \Eprint {https://arxiv.org/abs/1912.10010} {arXiv:1912.10010
  [hep-ph]} \BibitemShut {NoStop}%
\bibitem [{\citenamefont {Choubey}\ \emph {et~al.}(2021)\citenamefont
  {Choubey}, \citenamefont {Ghosh}, \citenamefont {Kempe},\ and\ \citenamefont
  {Ohlsson}}]{Choubey:2020dhw}%
  \BibitemOpen
  \bibfield  {author} {\bibinfo {author} {\bibfnamefont {S.}~\bibnamefont
  {Choubey}}, \bibinfo {author} {\bibfnamefont {M.}~\bibnamefont {Ghosh}},
  \bibinfo {author} {\bibfnamefont {D.}~\bibnamefont {Kempe}},\ and\ \bibinfo
  {author} {\bibfnamefont {T.}~\bibnamefont {Ohlsson}},\ }\href
  {https://doi.org/10.1007/JHEP05(2021)133} {\bibfield  {journal} {\bibinfo
  {journal} {JHEP}\ }\textbf {\bibinfo {volume} {05}},\ \bibinfo {pages}
  {133}},\ \Eprint {https://arxiv.org/abs/2010.16334} {arXiv:2010.16334
  [hep-ph]} \BibitemShut {NoStop}%
\bibitem [{\citenamefont {Majhi}\ \emph {et~al.}(2021)\citenamefont {Majhi},
  \citenamefont {Singha}, \citenamefont {Deepthi},\ and\ \citenamefont
  {Mohanta}}]{Majhi:2021api}%
  \BibitemOpen
  \bibfield  {author} {\bibinfo {author} {\bibfnamefont {R.}~\bibnamefont
  {Majhi}}, \bibinfo {author} {\bibfnamefont {D.~K.}\ \bibnamefont {Singha}},
  \bibinfo {author} {\bibfnamefont {K.~N.}\ \bibnamefont {Deepthi}},\ and\
  \bibinfo {author} {\bibfnamefont {R.}~\bibnamefont {Mohanta}},\ }\href
  {https://doi.org/10.1103/PhysRevD.104.055002} {\bibfield  {journal} {\bibinfo
   {journal} {Phys. Rev. D}\ }\textbf {\bibinfo {volume} {104}},\ \bibinfo
  {pages} {055002} (\bibinfo {year} {2021})},\ \Eprint
  {https://arxiv.org/abs/2101.08202} {arXiv:2101.08202 [hep-ph]} \BibitemShut
  {NoStop}%
\bibitem [{\citenamefont {Chatterjee}\ \emph {et~al.}(2022)\citenamefont
  {Chatterjee}, \citenamefont {Miranda}, \citenamefont {T\'ortola},\ and\
  \citenamefont {Valle}}]{Chatterjee:2021xyu}%
  \BibitemOpen
  \bibfield  {author} {\bibinfo {author} {\bibfnamefont {S.~S.}\ \bibnamefont
  {Chatterjee}}, \bibinfo {author} {\bibfnamefont {O.~G.}\ \bibnamefont
  {Miranda}}, \bibinfo {author} {\bibfnamefont {M.}~\bibnamefont {T\'ortola}},\
  and\ \bibinfo {author} {\bibfnamefont {J.~W.~F.}\ \bibnamefont {Valle}},\
  }\href {https://doi.org/10.1103/PhysRevD.106.075016} {\bibfield  {journal}
  {\bibinfo  {journal} {Phys. Rev. D}\ }\textbf {\bibinfo {volume} {106}},\
  \bibinfo {pages} {075016} (\bibinfo {year} {2022})},\ \Eprint
  {https://arxiv.org/abs/2111.08673} {arXiv:2111.08673 [hep-ph]} \BibitemShut
  {NoStop}%
\bibitem [{\citenamefont {Cordero}\ \emph {et~al.}(2023)\citenamefont
  {Cordero}, \citenamefont {Delgadillo},\ and\ \citenamefont
  {Miranda}}]{Cordero:2022fwb}%
  \BibitemOpen
  \bibfield  {author} {\bibinfo {author} {\bibfnamefont {R.}~\bibnamefont
  {Cordero}}, \bibinfo {author} {\bibfnamefont {L.~A.}\ \bibnamefont
  {Delgadillo}},\ and\ \bibinfo {author} {\bibfnamefont {O.~G.}\ \bibnamefont
  {Miranda}},\ }\href {https://doi.org/10.1103/PhysRevD.107.075023} {\bibfield
  {journal} {\bibinfo  {journal} {Phys. Rev. D}\ }\textbf {\bibinfo {volume}
  {107}},\ \bibinfo {pages} {075023} (\bibinfo {year} {2023})},\ \Eprint
  {https://arxiv.org/abs/2207.11308} {arXiv:2207.11308 [hep-ph]} \BibitemShut
  {NoStop}%
\bibitem [{\citenamefont {Cheng}\ \emph {et~al.}(2022)\citenamefont {Cheng},
  \citenamefont {Lindner},\ and\ \citenamefont {Rodejohann}}]{Cheng:2022lys}%
  \BibitemOpen
  \bibfield  {author} {\bibinfo {author} {\bibfnamefont {T.}~\bibnamefont
  {Cheng}}, \bibinfo {author} {\bibfnamefont {M.}~\bibnamefont {Lindner}},\
  and\ \bibinfo {author} {\bibfnamefont {W.}~\bibnamefont {Rodejohann}},\
  }\href {https://doi.org/10.1007/JHEP08(2022)111} {\bibfield  {journal}
  {\bibinfo  {journal} {JHEP}\ }\textbf {\bibinfo {volume} {08}},\ \bibinfo
  {pages} {111}},\ \Eprint {https://arxiv.org/abs/2204.10696} {arXiv:2204.10696
  [hep-ph]} \BibitemShut {NoStop}%
\bibitem [{\citenamefont {Aguilar}\ \emph
  {et~al.}(2024{\natexlab{b}})\citenamefont {Aguilar} \emph
  {et~al.}}]{ESSnuSB:2024yji}%
  \BibitemOpen
  \bibfield  {author} {\bibinfo {author} {\bibfnamefont {J.}~\bibnamefont
  {Aguilar}} \emph {et~al.} (\bibinfo {collaboration} {ESSnuSB}),\ }\href
  {https://doi.org/10.1007/JHEP08(2024)063} {\bibfield  {journal} {\bibinfo
  {journal} {JHEP}\ }\textbf {\bibinfo {volume} {08}},\ \bibinfo {pages}
  {063}},\ \Eprint {https://arxiv.org/abs/2404.17559} {arXiv:2404.17559
  [hep-ex]} \BibitemShut {NoStop}%
\bibitem [{\citenamefont {Aguilar}\ \emph {et~al.}(2025)\citenamefont {Aguilar}
  \emph {et~al.}}]{ESSnuSB:2025shd}%
  \BibitemOpen
  \bibfield  {author} {\bibinfo {author} {\bibfnamefont {J.}~\bibnamefont
  {Aguilar}} \emph {et~al.} (\bibinfo {collaboration} {ESSnuSB}),\ }\href
  {https://doi.org/10.1007/JHEP07(2025)186} {\bibfield  {journal} {\bibinfo
  {journal} {JHEP}\ }\textbf {\bibinfo {volume} {07}},\ \bibinfo {pages}
  {186}},\ \Eprint {https://arxiv.org/abs/2504.10480} {arXiv:2504.10480
  [hep-ph]} \BibitemShut {NoStop}%
\bibitem [{\citenamefont {Aguilar}\ \emph
  {et~al.}(2024{\natexlab{c}})\citenamefont {Aguilar} \emph
  {et~al.}}]{ESSnuSB:2024wet}%
  \BibitemOpen
  \bibfield  {author} {\bibinfo {author} {\bibfnamefont {J.}~\bibnamefont
  {Aguilar}} \emph {et~al.} (\bibinfo {collaboration} {ESSnuSB}),\ }\href
  {https://doi.org/10.1007/JHEP10(2024)187} {\bibfield  {journal} {\bibinfo
  {journal} {JHEP}\ }\textbf {\bibinfo {volume} {10}},\ \bibinfo {pages}
  {187}},\ \Eprint {https://arxiv.org/abs/2407.21663} {arXiv:2407.21663
  [hep-ex]} \BibitemShut {NoStop}%
\bibitem [{\citenamefont {Andreopoulos}\ \emph {et~al.}(2010)\citenamefont
  {Andreopoulos} \emph {et~al.}}]{Andreopoulos:2009rq}%
  \BibitemOpen
  \bibfield  {author} {\bibinfo {author} {\bibfnamefont {C.}~\bibnamefont
  {Andreopoulos}} \emph {et~al.} (\bibinfo {collaboration} {GENIE}),\ }\href
  {https://doi.org/10.1016/j.nima.2009.12.009} {\bibfield  {journal} {\bibinfo
  {journal} {Nucl. Instrum. Meth. A}\ }\textbf {\bibinfo {volume} {614}},\
  \bibinfo {pages} {87} (\bibinfo {year} {2010})},\ \Eprint
  {https://arxiv.org/abs/0905.2517} {arXiv:0905.2517 [hep-ph]} \BibitemShut
  {NoStop}%
\bibitem [{\citenamefont {Alvarez-Ruso}\ \emph {et~al.}(2021)\citenamefont
  {Alvarez-Ruso} \emph {et~al.}}]{GENIE:2021npt}%
  \BibitemOpen
  \bibfield  {author} {\bibinfo {author} {\bibfnamefont {L.}~\bibnamefont
  {Alvarez-Ruso}} \emph {et~al.} (\bibinfo {collaboration} {GENIE}),\ }\href
  {https://doi.org/10.1140/epjs/s11734-021-00295-7} {\bibfield  {journal}
  {\bibinfo  {journal} {Eur. Phys. J. ST}\ }\textbf {\bibinfo {volume} {230}},\
  \bibinfo {pages} {4449} (\bibinfo {year} {2021})},\ \Eprint
  {https://arxiv.org/abs/2106.09381} {arXiv:2106.09381 [hep-ph]} \BibitemShut
  {NoStop}%
\bibitem [{\citenamefont {Brun}\ and\ \citenamefont
  {Rademakers}(1997)}]{Brun:1997pa}%
  \BibitemOpen
  \bibfield  {author} {\bibinfo {author} {\bibfnamefont {R.}~\bibnamefont
  {Brun}}\ and\ \bibinfo {author} {\bibfnamefont {F.}~\bibnamefont
  {Rademakers}},\ }\href {https://doi.org/10.1016/S0168-9002(97)00048-X}
  {\bibfield  {journal} {\bibinfo  {journal} {Nucl. Instrum. Meth. A}\ }\textbf
  {\bibinfo {volume} {389}},\ \bibinfo {pages} {81} (\bibinfo {year}
  {1997})}\BibitemShut {NoStop}%
\bibitem [{\citenamefont {Huber}\ \emph {et~al.}(2005)\citenamefont {Huber},
  \citenamefont {Lindner},\ and\ \citenamefont {Winter}}]{Huber:2004ka}%
  \BibitemOpen
  \bibfield  {author} {\bibinfo {author} {\bibfnamefont {P.}~\bibnamefont
  {Huber}}, \bibinfo {author} {\bibfnamefont {M.}~\bibnamefont {Lindner}},\
  and\ \bibinfo {author} {\bibfnamefont {W.}~\bibnamefont {Winter}},\ }\href
  {https://doi.org/10.1016/j.cpc.2005.01.003} {\bibfield  {journal} {\bibinfo
  {journal} {Comput. Phys. Commun.}\ }\textbf {\bibinfo {volume} {167}},\
  \bibinfo {pages} {195} (\bibinfo {year} {2005})},\ \Eprint
  {https://arxiv.org/abs/hep-ph/0407333} {arXiv:hep-ph/0407333} \BibitemShut
  {NoStop}%
\bibitem [{\citenamefont {Huber}\ \emph {et~al.}(2007)\citenamefont {Huber},
  \citenamefont {Kopp}, \citenamefont {Lindner}, \citenamefont {Rolinec},\ and\
  \citenamefont {Winter}}]{Huber:2007ji}%
  \BibitemOpen
  \bibfield  {author} {\bibinfo {author} {\bibfnamefont {P.}~\bibnamefont
  {Huber}}, \bibinfo {author} {\bibfnamefont {J.}~\bibnamefont {Kopp}},
  \bibinfo {author} {\bibfnamefont {M.}~\bibnamefont {Lindner}}, \bibinfo
  {author} {\bibfnamefont {M.}~\bibnamefont {Rolinec}},\ and\ \bibinfo {author}
  {\bibfnamefont {W.}~\bibnamefont {Winter}},\ }\href
  {https://doi.org/10.1016/j.cpc.2007.05.004} {\bibfield  {journal} {\bibinfo
  {journal} {Comput. Phys. Commun.}\ }\textbf {\bibinfo {volume} {177}},\
  \bibinfo {pages} {432} (\bibinfo {year} {2007})},\ \Eprint
  {https://arxiv.org/abs/hep-ph/0701187} {arXiv:hep-ph/0701187} \BibitemShut
  {NoStop}%
\bibitem [{\citenamefont {Kopp}\ \emph {et~al.}(2008)\citenamefont {Kopp},
  \citenamefont {Ota},\ and\ \citenamefont {Winter}}]{Kopp:2008ds}%
  \BibitemOpen
  \bibfield  {author} {\bibinfo {author} {\bibfnamefont {J.}~\bibnamefont
  {Kopp}}, \bibinfo {author} {\bibfnamefont {T.}~\bibnamefont {Ota}},\ and\
  \bibinfo {author} {\bibfnamefont {W.}~\bibnamefont {Winter}},\ }\href
  {https://doi.org/10.1103/PhysRevD.78.053007} {\bibfield  {journal} {\bibinfo
  {journal} {Phys. Rev. D}\ }\textbf {\bibinfo {volume} {78}},\ \bibinfo
  {pages} {053007} (\bibinfo {year} {2008})},\ \Eprint
  {https://arxiv.org/abs/0804.2261} {arXiv:0804.2261 [hep-ph]} \BibitemShut
  {NoStop}%
\bibitem [{\citenamefont {Blennow}\ \emph {et~al.}(2020)\citenamefont
  {Blennow}, \citenamefont {Fernandez-Martinez}, \citenamefont {Ota},\ and\
  \citenamefont {Rosauro-Alcaraz}}]{Blennow:2019bvl}%
  \BibitemOpen
  \bibfield  {author} {\bibinfo {author} {\bibfnamefont {M.}~\bibnamefont
  {Blennow}}, \bibinfo {author} {\bibfnamefont {E.}~\bibnamefont
  {Fernandez-Martinez}}, \bibinfo {author} {\bibfnamefont {T.}~\bibnamefont
  {Ota}},\ and\ \bibinfo {author} {\bibfnamefont {S.}~\bibnamefont
  {Rosauro-Alcaraz}},\ }\href {https://doi.org/10.1140/epjc/s10052-020-7761-9}
  {\bibfield  {journal} {\bibinfo  {journal} {Eur. Phys. J. C}\ }\textbf
  {\bibinfo {volume} {80}},\ \bibinfo {pages} {190} (\bibinfo {year} {2020})},\
  \Eprint {https://arxiv.org/abs/1912.04309} {arXiv:1912.04309 [hep-ph]}
  \BibitemShut {NoStop}%
\bibitem [{\citenamefont {Ribeiro}\ \emph {et~al.}(2007)\citenamefont
  {Ribeiro}, \citenamefont {Minakata}, \citenamefont {Nunokawa}, \citenamefont
  {Uchinami},\ and\ \citenamefont {Zukanovich-Funchal}}]{Ribeiro:2007ud}%
  \BibitemOpen
  \bibfield  {author} {\bibinfo {author} {\bibfnamefont {N.~C.}\ \bibnamefont
  {Ribeiro}}, \bibinfo {author} {\bibfnamefont {H.}~\bibnamefont {Minakata}},
  \bibinfo {author} {\bibfnamefont {H.}~\bibnamefont {Nunokawa}}, \bibinfo
  {author} {\bibfnamefont {S.}~\bibnamefont {Uchinami}},\ and\ \bibinfo
  {author} {\bibfnamefont {R.}~\bibnamefont {Zukanovich-Funchal}},\ }\href
  {https://doi.org/10.1088/1126-6708/2007/12/002} {\bibfield  {journal}
  {\bibinfo  {journal} {JHEP}\ }\textbf {\bibinfo {volume} {12}},\ \bibinfo
  {pages} {002}},\ \Eprint {https://arxiv.org/abs/0709.1980} {arXiv:0709.1980
  [hep-ph]} \BibitemShut {NoStop}%
\bibitem [{\citenamefont {Choubey}\ and\ \citenamefont
  {Ohlsson}(2014)}]{Choubey:2014iia}%
  \BibitemOpen
  \bibfield  {author} {\bibinfo {author} {\bibfnamefont {S.}~\bibnamefont
  {Choubey}}\ and\ \bibinfo {author} {\bibfnamefont {T.}~\bibnamefont
  {Ohlsson}},\ }\href {https://doi.org/10.1016/j.physletb.2014.11.010}
  {\bibfield  {journal} {\bibinfo  {journal} {Phys. Lett. B}\ }\textbf
  {\bibinfo {volume} {739}},\ \bibinfo {pages} {357} (\bibinfo {year}
  {2014})},\ \Eprint {https://arxiv.org/abs/1410.0410} {arXiv:1410.0410
  [hep-ph]} \BibitemShut {NoStop}%
\bibitem [{\citenamefont {Aiello}\ \emph {et~al.}(2025)\citenamefont {Aiello}
  \emph {et~al.}}]{KM3NeT:2024pte}%
  \BibitemOpen
  \bibfield  {author} {\bibinfo {author} {\bibfnamefont {S.}~\bibnamefont
  {Aiello}} \emph {et~al.} (\bibinfo {collaboration} {KM3NeT}),\ }\href
  {https://doi.org/10.1088/1475-7516/2025/02/073} {\bibfield  {journal}
  {\bibinfo  {journal} {JCAP}\ }\textbf {\bibinfo {volume} {02}},\ \bibinfo
  {pages} {073}},\ \Eprint {https://arxiv.org/abs/2411.19078} {arXiv:2411.19078
  [hep-ex]} \BibitemShut {NoStop}%
\bibitem [{\citenamefont {Abbasi}\ \emph {et~al.}(2022)\citenamefont {Abbasi}
  \emph {et~al.}}]{IceCube:2022ubv}%
  \BibitemOpen
  \bibfield  {author} {\bibinfo {author} {\bibfnamefont {R.}~\bibnamefont
  {Abbasi}} \emph {et~al.} (\bibinfo {collaboration} {IceCube}),\ }\href
  {https://doi.org/10.1103/PhysRevLett.129.011804} {\bibfield  {journal}
  {\bibinfo  {journal} {Phys. Rev. Lett.}\ }\textbf {\bibinfo {volume} {129}},\
  \bibinfo {pages} {011804} (\bibinfo {year} {2022})},\ \Eprint
  {https://arxiv.org/abs/2201.03566} {arXiv:2201.03566 [hep-ex]} \BibitemShut
  {NoStop}%
\bibitem [{\citenamefont {Abbasi}\ \emph {et~al.}(2021)\citenamefont {Abbasi}
  \emph {et~al.}}]{IceCubeCollaboration:2021euf}%
  \BibitemOpen
  \bibfield  {author} {\bibinfo {author} {\bibfnamefont {R.}~\bibnamefont
  {Abbasi}} \emph {et~al.} (\bibinfo {collaboration} {IceCube}),\ }\href
  {https://doi.org/10.1103/PhysRevD.104.072006} {\bibfield  {journal} {\bibinfo
   {journal} {Phys. Rev. D}\ }\textbf {\bibinfo {volume} {104}},\ \bibinfo
  {pages} {072006} (\bibinfo {year} {2021})},\ \Eprint
  {https://arxiv.org/abs/2106.07755} {arXiv:2106.07755 [hep-ex]} \BibitemShut
  {NoStop}%
\bibitem [{\citenamefont {Albert}\ \emph {et~al.}(2022)\citenamefont {Albert}
  \emph {et~al.}}]{ANTARES:2021crm}%
  \BibitemOpen
  \bibfield  {author} {\bibinfo {author} {\bibfnamefont {A.}~\bibnamefont
  {Albert}} \emph {et~al.} (\bibinfo {collaboration} {ANTARES}),\ }\href
  {https://doi.org/10.1007/JHEP07(2022)048} {\bibfield  {journal} {\bibinfo
  {journal} {JHEP}\ }\textbf {\bibinfo {volume} {07}},\ \bibinfo {pages}
  {048}},\ \Eprint {https://arxiv.org/abs/2112.14517} {arXiv:2112.14517
  [hep-ex]} \BibitemShut {NoStop}%
\bibitem [{\citenamefont {Mitsuka}\ \emph {et~al.}(2011)\citenamefont {Mitsuka}
  \emph {et~al.}}]{Super-Kamiokande:2011dam}%
  \BibitemOpen
  \bibfield  {author} {\bibinfo {author} {\bibfnamefont {G.}~\bibnamefont
  {Mitsuka}} \emph {et~al.} (\bibinfo {collaboration} {Super-Kamiokande}),\
  }\href {https://doi.org/10.1103/PhysRevD.84.113008} {\bibfield  {journal}
  {\bibinfo  {journal} {Phys. Rev. D}\ }\textbf {\bibinfo {volume} {84}},\
  \bibinfo {pages} {113008} (\bibinfo {year} {2011})},\ \Eprint
  {https://arxiv.org/abs/1109.1889} {arXiv:1109.1889 [hep-ex]} \BibitemShut
  {NoStop}%
\bibitem [{\citenamefont {Coloma}\ \emph {et~al.}(2023)\citenamefont {Coloma},
  \citenamefont {Gonzalez-Garcia}, \citenamefont {Maltoni}, \citenamefont
  {Pinheiro},\ and\ \citenamefont {Urrea}}]{Coloma:2023ixt}%
  \BibitemOpen
  \bibfield  {author} {\bibinfo {author} {\bibfnamefont {P.}~\bibnamefont
  {Coloma}}, \bibinfo {author} {\bibfnamefont {M.~C.}\ \bibnamefont
  {Gonzalez-Garcia}}, \bibinfo {author} {\bibfnamefont {M.}~\bibnamefont
  {Maltoni}}, \bibinfo {author} {\bibfnamefont {J.~a.~P.}\ \bibnamefont
  {Pinheiro}},\ and\ \bibinfo {author} {\bibfnamefont {S.}~\bibnamefont
  {Urrea}},\ }\href {https://doi.org/10.1007/JHEP08(2023)032} {\bibfield
  {journal} {\bibinfo  {journal} {JHEP}\ }\textbf {\bibinfo {volume} {08}},\
  \bibinfo {pages} {032}},\ \Eprint {https://arxiv.org/abs/2305.07698}
  {arXiv:2305.07698 [hep-ph]} \BibitemShut {NoStop}%
\bibitem [{\citenamefont {Dziewonski}\ and\ \citenamefont
  {Anderson}(1981)}]{Dziewonski:1981xy}%
  \BibitemOpen
  \bibfield  {author} {\bibinfo {author} {\bibfnamefont {A.~M.}\ \bibnamefont
  {Dziewonski}}\ and\ \bibinfo {author} {\bibfnamefont {D.~L.}\ \bibnamefont
  {Anderson}},\ }\href {https://doi.org/10.1016/0031-9201(81)90046-7}
  {\bibfield  {journal} {\bibinfo  {journal} {Phys. Earth Planet. Interiors}\
  }\textbf {\bibinfo {volume} {25}},\ \bibinfo {pages} {297} (\bibinfo {year}
  {1981})}\BibitemShut {NoStop}%
\bibitem [{\citenamefont {Esteban}\ \emph
  {et~al.}(2024{\natexlab{b}})\citenamefont {Esteban} \emph
  {et~al.}}]{NuFIT:6.0}%
  \BibitemOpen
  \bibfield  {author} {\bibinfo {author} {\bibfnamefont {I.}~\bibnamefont
  {Esteban}} \emph {et~al.} (\bibinfo {collaboration} {NuFIT}),\ }\href@noop {}
  {\bibinfo {title} {Nufit 6.0}},\ \bibinfo {howpublished}
  {\url{http://www.nu-fit.org/}} (\bibinfo {year}
  {2024}{\natexlab{b}})\BibitemShut {NoStop}%
\bibitem [{\citenamefont {Honda}\ \emph {et~al.}(2015)\citenamefont {Honda},
  \citenamefont {Sajjad~Athar}, \citenamefont {Kajita}, \citenamefont
  {Kasahara},\ and\ \citenamefont {Midorikawa}}]{Honda:2015fha}%
  \BibitemOpen
  \bibfield  {author} {\bibinfo {author} {\bibfnamefont {M.}~\bibnamefont
  {Honda}}, \bibinfo {author} {\bibfnamefont {M.}~\bibnamefont {Sajjad~Athar}},
  \bibinfo {author} {\bibfnamefont {T.}~\bibnamefont {Kajita}}, \bibinfo
  {author} {\bibfnamefont {K.}~\bibnamefont {Kasahara}},\ and\ \bibinfo
  {author} {\bibfnamefont {S.}~\bibnamefont {Midorikawa}},\ }\href
  {https://doi.org/10.1103/PhysRevD.92.023004} {\bibfield  {journal} {\bibinfo
  {journal} {Phys. Rev. D}\ }\textbf {\bibinfo {volume} {92}},\ \bibinfo
  {pages} {023004} (\bibinfo {year} {2015})},\ \Eprint
  {https://arxiv.org/abs/1502.03916} {arXiv:1502.03916 [astro-ph.HE]}
  \BibitemShut {NoStop}%
\bibitem [{\citenamefont {Fogli}\ \emph {et~al.}(2002)\citenamefont {Fogli},
  \citenamefont {Lisi}, \citenamefont {Marrone}, \citenamefont {Montanino},\
  and\ \citenamefont {Palazzo}}]{Fogli:2002pt}%
  \BibitemOpen
  \bibfield  {author} {\bibinfo {author} {\bibfnamefont {G.~L.}\ \bibnamefont
  {Fogli}}, \bibinfo {author} {\bibfnamefont {E.}~\bibnamefont {Lisi}},
  \bibinfo {author} {\bibfnamefont {A.}~\bibnamefont {Marrone}}, \bibinfo
  {author} {\bibfnamefont {D.}~\bibnamefont {Montanino}},\ and\ \bibinfo
  {author} {\bibfnamefont {A.}~\bibnamefont {Palazzo}},\ }\href
  {https://doi.org/10.1103/PhysRevD.66.053010} {\bibfield  {journal} {\bibinfo
  {journal} {Phys. Rev. D}\ }\textbf {\bibinfo {volume} {66}},\ \bibinfo
  {pages} {053010} (\bibinfo {year} {2002})},\ \Eprint
  {https://arxiv.org/abs/hep-ph/0206162} {arXiv:hep-ph/0206162} \BibitemShut
  {NoStop}%
\bibitem [{\citenamefont {Gonzalez-Garcia}\ and\ \citenamefont
  {Maltoni}(2004)}]{Gonzalez-Garcia:2004pka}%
  \BibitemOpen
  \bibfield  {author} {\bibinfo {author} {\bibfnamefont {M.~C.}\ \bibnamefont
  {Gonzalez-Garcia}}\ and\ \bibinfo {author} {\bibfnamefont {M.}~\bibnamefont
  {Maltoni}},\ }\href {https://doi.org/10.1103/PhysRevD.70.033010} {\bibfield
  {journal} {\bibinfo  {journal} {Phys. Rev. D}\ }\textbf {\bibinfo {volume}
  {70}},\ \bibinfo {pages} {033010} (\bibinfo {year} {2004})},\ \Eprint
  {https://arxiv.org/abs/hep-ph/0404085} {arXiv:hep-ph/0404085} \BibitemShut
  {NoStop}%
\bibitem [{\citenamefont {Gandhi}\ \emph {et~al.}(2007)\citenamefont {Gandhi},
  \citenamefont {Ghoshal}, \citenamefont {Goswami}, \citenamefont {Mehta},
  \citenamefont {Sankar},\ and\ \citenamefont {Shalgar}}]{Gandhi:2007td}%
  \BibitemOpen
  \bibfield  {author} {\bibinfo {author} {\bibfnamefont {R.}~\bibnamefont
  {Gandhi}}, \bibinfo {author} {\bibfnamefont {P.}~\bibnamefont {Ghoshal}},
  \bibinfo {author} {\bibfnamefont {S.}~\bibnamefont {Goswami}}, \bibinfo
  {author} {\bibfnamefont {P.}~\bibnamefont {Mehta}}, \bibinfo {author}
  {\bibfnamefont {S.~U.}\ \bibnamefont {Sankar}},\ and\ \bibinfo {author}
  {\bibfnamefont {S.}~\bibnamefont {Shalgar}},\ }\href
  {https://doi.org/10.1103/PhysRevD.76.073012} {\bibfield  {journal} {\bibinfo
  {journal} {Phys. Rev. D}\ }\textbf {\bibinfo {volume} {76}},\ \bibinfo
  {pages} {073012} (\bibinfo {year} {2007})},\ \Eprint
  {https://arxiv.org/abs/0707.1723} {arXiv:0707.1723 [hep-ph]} \BibitemShut
  {NoStop}%
\bibitem [{\citenamefont {Ghosh}\ \emph {et~al.}(2013)\citenamefont {Ghosh},
  \citenamefont {Thakore},\ and\ \citenamefont {Choubey}}]{Ghosh:2012px}%
  \BibitemOpen
  \bibfield  {author} {\bibinfo {author} {\bibfnamefont {A.}~\bibnamefont
  {Ghosh}}, \bibinfo {author} {\bibfnamefont {T.}~\bibnamefont {Thakore}},\
  and\ \bibinfo {author} {\bibfnamefont {S.}~\bibnamefont {Choubey}},\ }\href
  {https://doi.org/10.1007/JHEP04(2013)009} {\bibfield  {journal} {\bibinfo
  {journal} {JHEP}\ }\textbf {\bibinfo {volume} {04}},\ \bibinfo {pages}
  {009}},\ \Eprint {https://arxiv.org/abs/1212.1305} {arXiv:1212.1305 [hep-ph]}
  \BibitemShut {NoStop}%
\bibitem [{\citenamefont {Choubey}\ \emph {et~al.}(2015)\citenamefont
  {Choubey}, \citenamefont {Ghosh}, \citenamefont {Ohlsson},\ and\
  \citenamefont {Tiwari}}]{Choubey:2015xha}%
  \BibitemOpen
  \bibfield  {author} {\bibinfo {author} {\bibfnamefont {S.}~\bibnamefont
  {Choubey}}, \bibinfo {author} {\bibfnamefont {A.}~\bibnamefont {Ghosh}},
  \bibinfo {author} {\bibfnamefont {T.}~\bibnamefont {Ohlsson}},\ and\ \bibinfo
  {author} {\bibfnamefont {D.}~\bibnamefont {Tiwari}},\ }\href
  {https://doi.org/10.1007/JHEP12(2015)126} {\bibfield  {journal} {\bibinfo
  {journal} {JHEP}\ }\textbf {\bibinfo {volume} {12}},\ \bibinfo {pages}
  {126}},\ \Eprint {https://arxiv.org/abs/1507.02211} {arXiv:1507.02211
  [hep-ph]} \BibitemShut {NoStop}%
\bibitem [{\citenamefont {Fukasawa}\ and\ \citenamefont
  {Yasuda}(2015)}]{Fukasawa:2015jaa}%
  \BibitemOpen
  \bibfield  {author} {\bibinfo {author} {\bibfnamefont {S.}~\bibnamefont
  {Fukasawa}}\ and\ \bibinfo {author} {\bibfnamefont {O.}~\bibnamefont
  {Yasuda}},\ }\href {https://doi.org/10.1155/2015/820941} {\bibfield
  {journal} {\bibinfo  {journal} {Adv. High Energy Phys.}\ }\textbf {\bibinfo
  {volume} {2015}},\ \bibinfo {pages} {820941} (\bibinfo {year} {2015})},\
  \Eprint {https://arxiv.org/abs/1503.08056} {arXiv:1503.08056 [hep-ph]}
  \BibitemShut {NoStop}%
\bibitem [{\citenamefont {Fukasawa}\ and\ \citenamefont
  {Yasuda}(2017)}]{Fukasawa:2016nwn}%
  \BibitemOpen
  \bibfield  {author} {\bibinfo {author} {\bibfnamefont {S.}~\bibnamefont
  {Fukasawa}}\ and\ \bibinfo {author} {\bibfnamefont {O.}~\bibnamefont
  {Yasuda}},\ }\href {https://doi.org/10.1016/j.nuclphysb.2016.11.004}
  {\bibfield  {journal} {\bibinfo  {journal} {Nucl. Phys. B}\ }\textbf
  {\bibinfo {volume} {914}},\ \bibinfo {pages} {99} (\bibinfo {year} {2017})},\
  \Eprint {https://arxiv.org/abs/1608.05897} {arXiv:1608.05897 [hep-ph]}
  \BibitemShut {NoStop}%
\bibitem [{\citenamefont {Kelly}(2017)}]{Kelly:2017kch}%
  \BibitemOpen
  \bibfield  {author} {\bibinfo {author} {\bibfnamefont {K.~J.}\ \bibnamefont
  {Kelly}},\ }\href {https://doi.org/10.1103/PhysRevD.95.115009} {\bibfield
  {journal} {\bibinfo  {journal} {Phys. Rev. D}\ }\textbf {\bibinfo {volume}
  {95}},\ \bibinfo {pages} {115009} (\bibinfo {year} {2017})},\ \Eprint
  {https://arxiv.org/abs/1703.00448} {arXiv:1703.00448 [hep-ph]} \BibitemShut
  {NoStop}%
\bibitem [{\citenamefont {Choubey}(2016)}]{Choubey:2016gps}%
  \BibitemOpen
  \bibfield  {author} {\bibinfo {author} {\bibfnamefont {S.}~\bibnamefont
  {Choubey}},\ }\href {https://doi.org/10.1016/j.nuclphysb.2016.03.026}
  {\bibfield  {journal} {\bibinfo  {journal} {Nucl. Phys. B}\ }\textbf
  {\bibinfo {volume} {908}},\ \bibinfo {pages} {235} (\bibinfo {year}
  {2016})},\ \Eprint {https://arxiv.org/abs/1603.06841} {arXiv:1603.06841
  [hep-ph]} \BibitemShut {NoStop}%
\bibitem [{\citenamefont {An}\ \emph {et~al.}(2016)\citenamefont {An} \emph
  {et~al.}}]{JUNO:2015zny}%
  \BibitemOpen
  \bibfield  {author} {\bibinfo {author} {\bibfnamefont {F.}~\bibnamefont {An}}
  \emph {et~al.} (\bibinfo {collaboration} {JUNO}),\ }\href@noop {} {\bibfield
  {journal} {\bibinfo  {journal} {J. Phys. G}\ }\textbf {\bibinfo {volume}
  {43}},\ \bibinfo {pages} {030401} (\bibinfo {year} {2016})},\ \Eprint
  {https://arxiv.org/abs/1507.05613} {arXiv:1507.05613 [physics.ins-det]}
  \BibitemShut {NoStop}%
\bibitem [{\citenamefont {Abi}\ \emph {et~al.}(2020)\citenamefont {Abi} \emph
  {et~al.}}]{DUNE:2020ypp}%
  \BibitemOpen
  \bibfield  {author} {\bibinfo {author} {\bibfnamefont {B.}~\bibnamefont
  {Abi}} \emph {et~al.} (\bibinfo {collaboration} {DUNE}),\ }\href@noop {} {\
  (\bibinfo {year} {2020})},\ \Eprint {https://arxiv.org/abs/2002.03005}
  {arXiv:2002.03005 [hep-ex]} \BibitemShut {NoStop}%
\bibitem [{\citenamefont {Aartsen}\ \emph {et~al.}(2021)\citenamefont {Aartsen}
  \emph {et~al.}}]{IceCube-Gen2:2020qha}%
  \BibitemOpen
  \bibfield  {author} {\bibinfo {author} {\bibfnamefont {M.~G.}\ \bibnamefont
  {Aartsen}} \emph {et~al.} (\bibinfo {collaboration} {IceCube-Gen2}),\ }\href
  {https://doi.org/10.1088/1361-6471/abbd48} {\bibfield  {journal} {\bibinfo
  {journal} {J. Phys. G}\ }\textbf {\bibinfo {volume} {48}},\ \bibinfo {pages}
  {060501} (\bibinfo {year} {2021})},\ \Eprint
  {https://arxiv.org/abs/2008.04323} {arXiv:2008.04323 [astro-ph.HE]}
  \BibitemShut {NoStop}%
\bibitem [{\citenamefont {Adrian-Martinez}\ \emph {et~al.}(2016)\citenamefont
  {Adrian-Martinez} \emph {et~al.}}]{KM3Net:2016zxf}%
  \BibitemOpen
  \bibfield  {author} {\bibinfo {author} {\bibfnamefont {S.}~\bibnamefont
  {Adrian-Martinez}} \emph {et~al.} (\bibinfo {collaboration} {KM3Net}),\
  }\href {https://doi.org/10.1088/0954-3899/43/8/084001} {\bibfield  {journal}
  {\bibinfo  {journal} {J. Phys. G}\ }\textbf {\bibinfo {volume} {43}},\
  \bibinfo {pages} {084001} (\bibinfo {year} {2016})},\ \Eprint
  {https://arxiv.org/abs/1601.07459} {arXiv:1601.07459 [astro-ph.IM]}
  \BibitemShut {NoStop}%
\bibitem [{\citenamefont {Agostini}\ \emph {et~al.}(2020)\citenamefont
  {Agostini} \emph {et~al.}}]{P-ONE:2020ljt}%
  \BibitemOpen
  \bibfield  {author} {\bibinfo {author} {\bibfnamefont {M.}~\bibnamefont
  {Agostini}} \emph {et~al.} (\bibinfo {collaboration} {P-ONE}),\ }\href
  {https://doi.org/10.1038/s41550-020-1182-4} {\bibfield  {journal} {\bibinfo
  {journal} {Nature Astron.}\ }\textbf {\bibinfo {volume} {4}},\ \bibinfo
  {pages} {913} (\bibinfo {year} {2020})},\ \Eprint
  {https://arxiv.org/abs/2005.09493} {arXiv:2005.09493 [astro-ph.HE]}
  \BibitemShut {NoStop}%
\bibitem [{\citenamefont {Chatterjee}\ and\ \citenamefont
  {De~Roeck}(2024)}]{Chatterjee:2024ein}%
  \BibitemOpen
  \bibfield  {author} {\bibinfo {author} {\bibfnamefont {A.}~\bibnamefont
  {Chatterjee}}\ and\ \bibinfo {author} {\bibfnamefont {A.}~\bibnamefont
  {De~Roeck}},\ }\href {https://doi.org/10.1016/j.physletb.2024.138838}
  {\bibfield  {journal} {\bibinfo  {journal} {Phys. Lett. B}\ }\textbf
  {\bibinfo {volume} {855}},\ \bibinfo {pages} {138838} (\bibinfo {year}
  {2024})},\ \Eprint {https://arxiv.org/abs/2402.16441} {arXiv:2402.16441
  [hep-ph]} \BibitemShut {NoStop}%
\end{thebibliography}%
\end{document}